\documentclass[%
 reprint,
 superscriptaddress,
 longbibliography,
 amsmath,amssymb,
 aps,
 prx,
floatfix,
]{revtex4-2}
\usepackage[english]{babel}
\usepackage{graphicx}% Include figure files
\usepackage{dcolumn}% Align table columns on decimal point
\usepackage{bm}% bold math
\usepackage{comment}
\usepackage{xcolor}
\usepackage{amsmath}
\usepackage{braket}
\usepackage[normalem]{ulem}
\usepackage{float}
\usepackage{censor}
\usepackage[colorlinks,citecolor=red,urlcolor=blue,bookmarks=false,hypertexnames=true]{hyperref}

\usepackage[autostyle]{csquotes}
\MakeOuterQuote{"}
\usepackage{todonotes}

\begin{document}

\preprint{APS/123-QED}

\title{A Self-Similar Sine-Cosine Fractal Architecture for Multiport Interferometers}% Force line breaks with \\

\author{Jasvith Raj Basani}%
\affiliation{Department of Electrical and Computer Engineering, Institute for Research in Electronics and Applied Physics, and Joint Quantum Institute, University of Maryland, College Park, MD 20742, USA}

\author{Sri Krishna Vadlamani}%
\affiliation{ Research Laboratory of Electronics, Massachusetts Institute of Technology, 77 Massachusetts Avenue, Cambridge, Massachusetts 02139, USA}%

\author{Saumil Bandyopadhyay}%
\affiliation{ Research Laboratory of Electronics, Massachusetts Institute of Technology, 77 Massachusetts Avenue, Cambridge, Massachusetts 02139, USA}%

\author{Dirk R. Englund}%
\affiliation{ Research Laboratory of Electronics, Massachusetts Institute of Technology,
77 Massachusetts Avenue, Cambridge, Massachusetts 02139, USA}%

\author{Ryan Hamerly}%
\affiliation{ Research Laboratory of Electronics, Massachusetts Institute of Technology,
77 Massachusetts Avenue, Cambridge, Massachusetts 02139, USA}%
\affiliation{NTT Research Inc., PHI Laboratories, 940 Stewart Drive, Sunnyvale, CA 94085, USA}

\date{\today}% It is always \today, today,
             %  but any date may be explicitly specified

\begin{abstract}
Multiport interferometers based on integrated beamsplitter meshes have recently captured interest as a platform for many emerging technologies. In this paper, we present a novel architecture for multiport interferometers based on the Sine-Cosine fractal decomposition of a unitary matrix. Our architecture is unique in that it is self-similar, enabling the construction of modular multi-chiplet devices. Due to this modularity, our design enjoys improved resilience to hardware imperfections as compared to conventional multiport interferometers. Additionally, the structure of our circuit enables systematic truncation, which is key in reducing the hardware footprint of the chip as well as compute time in training optical neural networks, while maintaining full connectivity. Numerical simulations show that truncation of these meshes gives robust performance even under large fabrication errors. This design is a step forward in the construction of large-scale programmable photonics, removing a major hurdle in scaling up to practical machine learning and quantum computing applications.
\end{abstract}

%\keywords{Suggested keywords}%Use showkeys class option if keyword
                              %display desired
\maketitle

%\tableofcontents

\section{Introduction}

Photonic Integrated Circuits (PICs) have recently captured interest as a promising time- and energy-efficient platform for classical and quantum optical information processing. They have been used to accelerate tasks in signal processing \cite{annoni2017unscrambling, ribeiro2016demonstration, milanizadeh2019manipulating, zhuang2015programmable, notaros2017programmable}, machine learning \cite{NN_arch_1, opt_ml}, optimization \cite{prabhu2020ising}, and quantum simulation \cite{harris2017quantum, wang2018multidimensional, qiang2018large, sparrow2018simulating, carolan2015universal}. Scaling these systems up in order to tackle real-world problems requires careful attention to issues such as the effect of analog component imperfections on performance, and the scaling of chip area with system size. 

For instance, it has been shown that the test accuracy of optical neural networks (ONNs) based on Mach-Zehnder Interferometer (MZI) meshes \cite{NN_arch_1} drops rapidly as soon as the constituent beamsplitters deviate from 50-50 splitting ratio by a couple of percent \cite{NN_arch_2}. A variety of error-correction techniques have been proposed for the MZI-based platform\textemdash global optimization \cite{imperfect_mesh_1, imperfect_mesh_2, lopez2019programmable, lopez2020auto, perez2020multipurpose, imperfect_mesh_3}, local correction \cite{hec_1, imperfect_mesh_4}, self-configuration \cite{imperfect_mesh_5, self_con_1, self_con_2, NN_arch_3}, and hardware augmentation \cite{Inf, suzuki2015ultra, miller2015perfect}. The behavior of many of these techniques can be better understood by considering an important insight derived in Ref.~\cite{hec_1}\textemdash MZIs with imperfect beamsplitters implement only a subset of all the $2\times2$ unitary matrices that a perfect MZI can implement. This fact explains the observed imperfection-induced reduction in ONN performance since circuits composed of imperfect MZIs implement fewer functions than those with perfect MZIs \cite{NN_arch_1, NN_arch_2}.

We show in this paper that the extent of reduction of the expressivity of a faulty MZI mesh depends strongly on its geometry and that a careful choice of mesh geometry can significantly soften the negative impact of hardware errors. We do so by introducing a novel self-similar MZI-mesh architecture based on the recursive Sine-Cosine unitary decomposition of Polcari \cite{butterfly_decomp} and demonstrating that it is more robust to MZI errors than the conventional \texttt{RECK} (triangular) \cite{reck} and \texttt{CLEMENTS} (rectangular) \cite{clements} mesh geometries. The recursive Sine-Cosine decomposition \cite{butterfly_decomp} is a generalization of the standard FFT decomposition of Fourier Transform matrices~\cite{butterfly_og} to arbitrary unitary matrices. We shall refer to MZI meshes constructed using this decomposition as Sine-Cosine Fractal (SCF) meshes. Like the FFT mesh~\cite{FFT_analog_1, FFT_analog_2}, the SCF mesh has a recursive, self-similar structure; the FFT mesh can in fact be obtained from the SCF mesh by the mere pruning (omission) of certain columns of MZIs. 

While SCF meshes have greater error robustness than other architectures, they can also be systematically shrunk in size for use in machine learning applications. In analogy with pruning in conventional neural networks~\cite{conventional_prune_1, conventional_prune_2}, we introduce a systematic mesh pruning scheme that interpolates between the simple FFT and the full Sine-Cosine Fractal, and numerically demonstrate that ONNs composed of pruned meshes still achieve excellent performance at benchmark learning tasks.  

% Generic global optimization can be time-consuming because it has to be applied separately to each individual faulty circuit while hardware augmentation increases the chip footprint.

The paper is organized as follows: Section~\ref{sec:2} introduces and discusses the Sine-Cosine Fractal architecture. Section~\ref{sec:3} contains analytical and numerical results on the expressivity of SCF meshes in the presence of beamsplitter errors. Section~\ref{sec:4} reports the performance of ONNs constructed from both complete and pruned SCF meshes, and Section~\ref{sec:5} concludes the paper with a further discussion of the scope and impact of our work.

\section{\label{sec:2}The Sine-Cosine Fractal Architecture}

\begin{figure}
    \centering
    \includegraphics[width = \columnwidth]{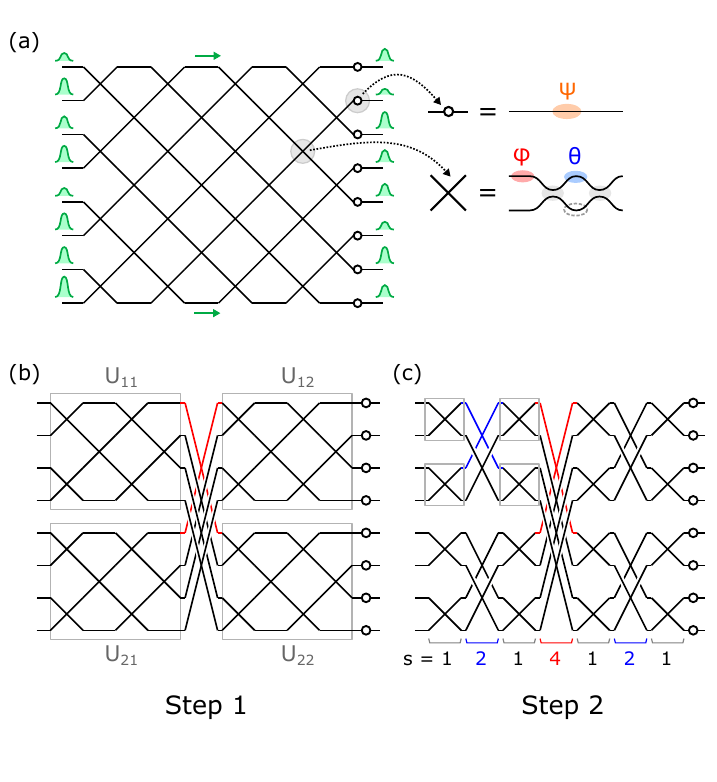}
    \caption{\textbf{(a)} $8 \times 8$ \texttt{CLEMENTS} mesh. \textbf{(b)} First step of the block decomposition of the mesh. This results in $4 \times 4$ quadrants. \textbf{(c)} Further decomposition of the quadrants, which results in the SCF mesh. }
    \label{fig_1}
\end{figure}

In order to construct a photonic circuit that implements a given unitary matrix $U$, one first decomposes $U$ into a product of $2\times 2$ unitary matrices and diagonal phase shifts. MZIs are used to implement the $2\times 2$ unitary matrices in the hardware; the transfer function of an MZI with two phase-shifters $\theta,\phi$ (Fig.~\ref{fig_1}(a)) is given by:
\begin{align}
	T(\theta,\phi) & = \frac{1}{2} \begin{bmatrix} 1 & i \\ i & 1 \end{bmatrix} \begin{bmatrix} e^{i\theta} & 0 \\ 0 & 1 \end{bmatrix}
				  	    \begin{bmatrix} 1 & i \\ i & 1 \end{bmatrix} \begin{bmatrix} e^{i\phi} & 0 \\ 0 & 1 \end{bmatrix}
					    \nonumber \\
	  & = i e^{i\theta/2} \begin{bmatrix} e^{i\phi} \sin(\theta/2) & \cos(\theta/2) \\ e^{i\phi} \cos(\theta/2) & \sin(\theta/2) \end{bmatrix} \label{eq:t}
\end{align}
The arrangement of MZIs in the circuit, that is, the geometry of the MZI mesh, determines the order of appearance of the corresponding $2\times 2$ unitaries in the decomposition. Fig.~\ref{fig_1}(a) depicts a mesh that implements an $8\times 8$ matrix via the \texttt{CLEMENTS} decomposition~\cite{clements}. The \texttt{RECK}~\cite{reck} and balanced binary tree~\cite{binary_tree_decomp} are other important decompositions that have been used to construct unitary meshes. 

This paper proposes a new mesh architecture based on the Sine-Cosine decomposition, a block diagonalization of unitary matrices, which on an $N\times N$ matrix $U$ consists of partitioning $U$ into four $N/2\times N/2$ blocks and performing the Singular Value Decomposition (SVD) on each block. The unitarity of $U$ imposes special constraints that force the blocks to share singular vectors. Polcari \cite{butterfly_decomp} shows that the block-wise SVD of $U$ yields:
\begin{equation}
    U = \begin{bmatrix} 
    U_{12} &0 \\ 
      0    & U_{22}
    \end{bmatrix}
    \begin{bmatrix}
    D_{11} & D_{12} \\ 
    D_{21} & D_{22}
    \end{bmatrix}
    \begin{bmatrix}
    U_{11} &0 \\
    0 & U_{21}
    \end{bmatrix}
    \label{eq:udiv}
\end{equation}
where $U_{11}, U_{12}, U_{21}, U_{22}$ are unitary matrices and $D_{11}, D_{12}, D_{21}, D_{22}$ are diagonal matrices encoding the singular vectors and values, respectively.  Unitarity constrains the $D_{ij}$ to take the following form: 
\begin{equation}
	\begin{bmatrix} D_{11} & D_{12} \\ D_{21} & D_{22} \end{bmatrix} = i e^{i\Theta/2} \begin{bmatrix} e^{i\Phi}\sin(\Theta/2) & \cos(\Theta/2) \\ e^{i\Phi}\cos(\Theta/2) & -\sin(\Theta/2) \end{bmatrix} \label{eq:dij}
\end{equation}
with $(\Theta, \Phi)$ representing diagonal matrices that encode phase shifts. Fig.~\ref{fig_1}(b) depicts Eq.~\eqref{eq:udiv} graphically for the $8\times8$ case\textemdash the $4\times4$ unitary matrices $U_{ij}$ are implemented by Clements meshes while the diagonal matrices $D_{ij}$ are implemented by MZIs in the center that couple the four unitary blocks. One can actually go further and perform the block-wise SVD again on each of the $U_{ij}$ subblocks to obtain the mesh of Fig.~\ref{fig_1}(c); the $2\times2$ unitaries obtained from the $4\times4$ unitaries are now directly implemented by MZIs. Because of its self-similar structure, we denote this geometry the sine-cosine fractal (SCF) mesh. In the general case, an SCF mesh can be constructed from any radix-2 ($N = 2^n$) matrix: one recursively performs blockwise SVDs on each unitary matrix of size greater than $2$ until the full decomposition consists only of $2\times2$ matrices that connect different modes.      

Like the \texttt{CLEMENTS} and other conventional architectures, the SCF mesh has a depth that scales as $O(N)$, has $N^{2}$ degrees of freedom, and is universal, i.e.\ it can represent the entire unitary group. However, the SCF mesh also possesses tunable crossings that couple non-neighboring waveguides (Fig.~\ref{fig_1}(c)), i.e.\ crossings of {\it stride} $s > 1$, interleaved between conventional crossings with $s = 1$.  This is in contrast to conventional architectures, where all MZIs have unit stride.

\section{\label{sec:3}Error Correction and Matrix Fidelity}

The long-stride crossings give the SCF mesh greater robustness to hardware imperfections.  To see how, consider the problem of realizing a target at high fidelity on imperfect hardware.  As mentioned previously, MZI meshes with perfect beamsplitters can implement any unitary matrix; however, the introduction of faults, however, reduces the expressivity of the mesh and consequently the fraction of matrices that are implementable drops below unity. In this section, we show that Sine-Cosine Fractal meshes can perfectly implement a greater fraction of random matrices than Clements meshes can for the same beamsplitter error level.  

\subsection{Distribution of mesh phase-shifts for Haar-random matrices}

\begin{figure}
    \centering
    \includegraphics[width = \columnwidth]{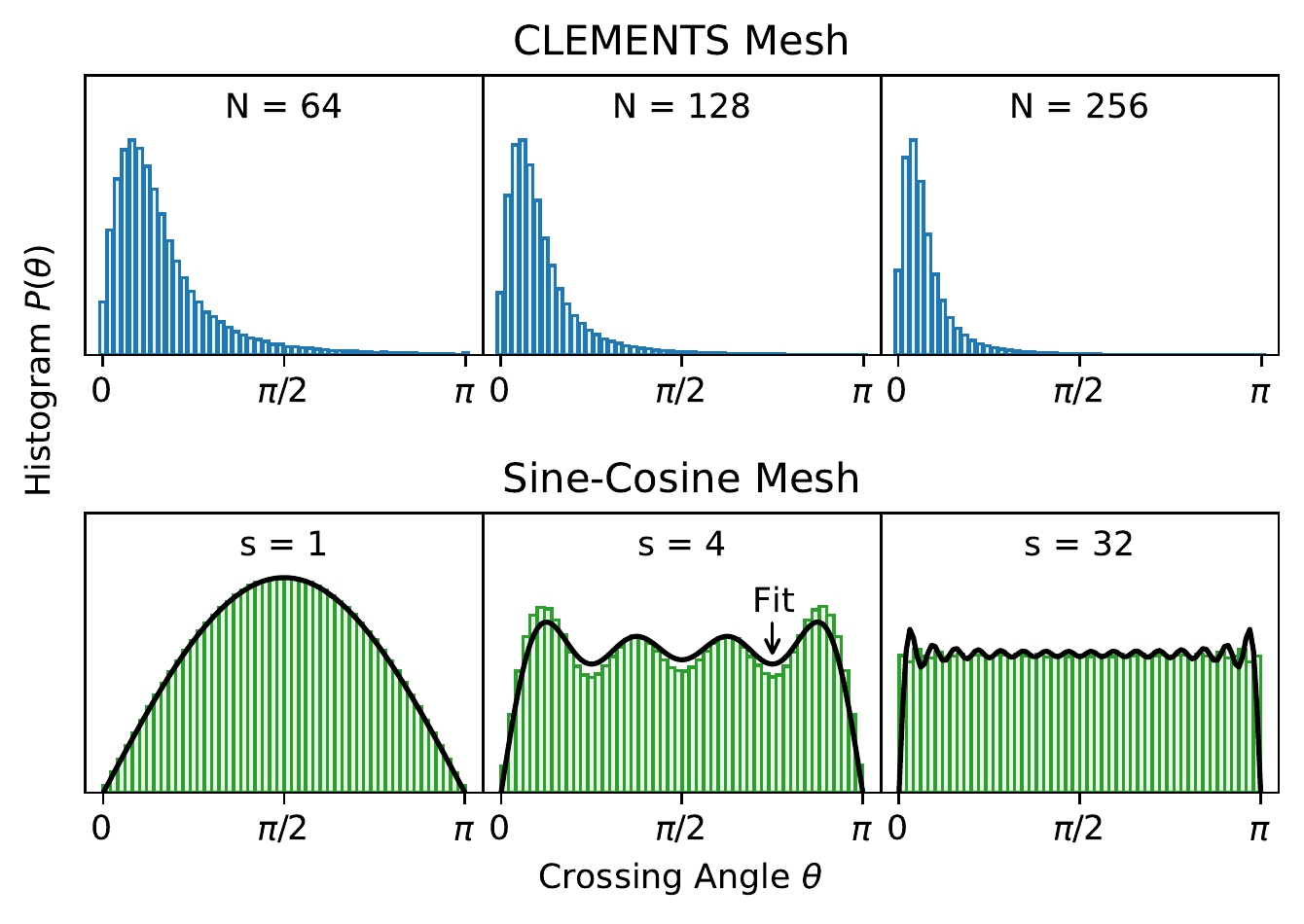}
    \caption{Distribution of crossing angles for the \texttt{CLEMENTS} and SCF meshes as a function of mesh size and stride respectively. The crossing angles of the \texttt{CLEMENTS} mesh increasingly clusters near $\theta = 0$ for larger meshes. Crossing angles for the SCF mesh become increasingly uniformly distributed as the stride increases.}
    \label{fig_2}
\end{figure}

A specific setting of phase shifts $(\theta, \phi)$ is required to make a mesh implement a given target unitary matrix. Drawing the target matrix from the Haar (uniform) distribution \cite{Haar_sampling} induces a distribution $P(\theta)$ over the phase-shifts. Russell et al. \cite{MZI_rank} show that the phase-shift distribution of the $n$-th MZI (according to any indexing scheme) in either the \texttt{RECK} or \texttt{CLEMENTS} meshes is given by $P_n(\theta) = k_n \mathrm{sin}(\theta/2) \mathrm{cos}(\theta/2)^{2k_n - 1}$, where $k_n \in \{1, \hdots, N - 1 \}$, called the \textit{rank} of the $n$-th MZI, is a function of the physical position of the MZI within the mesh. There are $(N - k)$ MZIs of rank $k$~\cite{MZI_rank}. For larger meshes, the rank of an average MZI increases, and this results in $P(\theta)$, which is an average of the $P_n(\theta)$ over all the MZIs, clustering around the cross state $\theta = 0$ (top row of Fig.~\ref{fig_2}). 

However, unlike the \texttt{RECK} and \texttt{CLEMENTS} meshes, the SCF mesh is configured from a top-down block decompsition of the matrix.  For a given matrix $U$, sampled over the Haar measure, the singular-vector matrices $U_{ij}, (i,j) \in \{1, 2\}$ in Eq.~(\ref{eq:udiv}) are also Haar-random and independent of each other. As a result, the distribution $P_n(\theta)$ depends only on the stride $s_n$ of the $n$-th MZI, not on its location in the mesh. The bottom row of Fig.~\ref{fig_2} shows the distribution of angles for MZIs of different stride.  The majority of MZIs have unit stride and $P_n(\theta) \propto \sin(\theta)$. As the stride increases, this distribution begins to resemble a uniform distribution. We find that $P_n(\theta)$ for stride $s_n$ can be fit well by the normalized finite Fourier series of a constant given by:
\begin{equation}
    	P_n(\theta) \approx \frac{1}{2\sum_{q=1}^{s_n}{(2q-1)^{-2}}} \sum_{q=1}^{s_n}{\frac{\sin\bigl((2q-1)\theta\bigr)}{2q-1}}\label{eq:scffourier}
\end{equation}

\subsection{Error in implementing Haar-random matrices in the presence of MZI errors}
The deviations from 50:50 of the two constituent beamsplitters of fabricated MZIs are captured by the phase angles ($\alpha, \beta$). These splitter errors perturb the MZI transfer matrix as follows: 
\begin{align}
	T'(\theta',\phi',\alpha,\beta) & = 
	\begin{bmatrix} \cos(\tfrac\pi4\!+\!\beta) & i\sin(\tfrac\pi4\!+\!\beta) \\ i\sin(\tfrac\pi4\!+\!\beta) & \cos(\tfrac\pi4\!+\!\beta) \end{bmatrix} \begin{bmatrix} e^{i\theta'} & 0 \\ 0 & 1 \end{bmatrix} \nonumber \\
	& \ \ \ \ \times 	\begin{bmatrix} \cos(\tfrac\pi4\!+\!\alpha) & i\sin(\tfrac\pi4\!+\!\alpha) \\ i\sin(\tfrac\pi4\!+\!\alpha) & \cos(\tfrac\pi4\!+\!\alpha) \end{bmatrix} \begin{bmatrix} e^{i\phi'} & 0 \\ 0 & 1 \end{bmatrix} \label{eq:dt}
\end{align}
Bandyopadhyay et al. \cite{hec_1} show that it is always possible to choose phase-shifts $\theta',\phi'$ for a faulty MZI with errors ($\alpha, \beta$) such that it implements the transfer matrix of an ideal MZI $T(\theta,\phi)$ as long as:
\begin{equation}
    \underbrace{2|\alpha+\beta|}_{\theta_{\rm min}} \leq \theta \leq \underbrace{\pi - 2|\alpha-\beta|}_{\theta_{\rm max}} \label{eq:thetarange}
\end{equation}
If $\theta$ is outside this range, the faulty MZI cannot exactly emulate the ideal MZI and the faulty mesh transfer function deviates from that of the ideal mesh which implements the target matrix.

For a given target unitary $U$, we quantify the deviation of the faulty mesh using the Frobenius norm $\mathcal{E} = \lVert\Delta U\rVert/\sqrt{N}$ which computes the average relative error per matrix element. This quantity is then averaged over both the choice of target unitary (from the Haar distribution), and the distribution of MZI splitter errors, which are assumed to be independent Gaussians ($\alpha, \beta \sim \mathcal{N}(0, \sigma)$). In the case of correlated errors, the effect of most correlations vanish over the Haar measure as proven in Ref.~\cite{self_con_1}. 

In the absence of error-correction techniques, all the MZIs of the ideal mesh are implemented incorrectly by the corresponding MZIs in the faulty mesh and the error Frobenius norm $\mathcal{E}_{0}$ (for both \texttt{CLEMENTS} and SCF meshes) has been shown~\cite{self_con_1} to scale linearly with the MZI error $\sigma$ as $\mathcal{E}_{0} = \sqrt{2N}\sigma$. When one uses the error-correction techniques of Refs.~\cite{hec_1, self_con_1, self_con_2}, only the MZIs in the ideal mesh that do not satisfy Eq.~\eqref{eq:thetarange} are implemented incorrectly by the corresponding MZIs in the faulty mesh\textemdash only those MZIs contribute to the error Frobenius norm. Using $n$ to denote the MZI location in the mesh as before, the average ``corrected'' error, $\mathcal{E}_{c}$ (assuming uncorrelated errors, see \cite[Supp.~Sec.~1]{Inf}), is computed by integrating over the probability that each MZI does not satisfy Eq.~\eqref{eq:thetarange}:
\begin{align}
    	(\mathcal{E}_c)^2 & = \frac{1}{2N}\sum_n \Bigl[\int_{0}^{\theta_{\rm min}} P_n(\theta)(\theta-\theta_{\rm min})^2\mathrm{d}\theta \nonumber \\
    		& \qquad\qquad\ \  + \int_{\theta_{\rm max}}^{\pi} P_n(\theta)(\theta-\theta_{\rm max})^2\mathrm{d}\theta\Bigr] 
    \label{eq:ec}
\end{align}
Eq.~\eqref{eq:ec} indicates that the effectiveness of error-correction, measured by $\mathcal{E}_c$, is strongly dependent on the distribution $P_{n}(\theta)$. For the \texttt{CLEMENTS} mesh, Ref.~\cite{self_con_1} proves that $\mathcal{E}_{c}^{\mathrm{(clem)}} = \sqrt{\frac{2}{3}} N\sigma^{2}$, which is a quadratic improvement over $\mathcal{E}_0$. 

Since the integral is over angles close to either $\theta=0$ or $\theta=\pi$, the two terms in Eq.~\eqref{eq:ec} are estimated for the Sine-Cosine Fractal mesh by Taylor-expanding Eq.~\eqref{eq:scffourier} to first order about $\theta=0,\pi$ respectively. The result is:
\begin{equation}
    \mathcal{E}_c^{\mathrm{(scf)}} = \frac{4}{\pi} \sqrt{N\log_2(N)}\, \sigma^2. \label{eq:ecft}
\end{equation}
The ratio of corrected errors for both meshes is:
\begin{equation}
    \frac{\mathcal{E}_c^{\rm (clem)}}{\mathcal{E}_c^{\rm (scf)}} = \sqrt{\frac{\pi^2}{24} \frac{N}{\log_2(N)}}
\end{equation}
which is greater than $1$ for all but very small $N$.

\begin{figure}[t]
    \centering
    \includegraphics[width = \columnwidth]{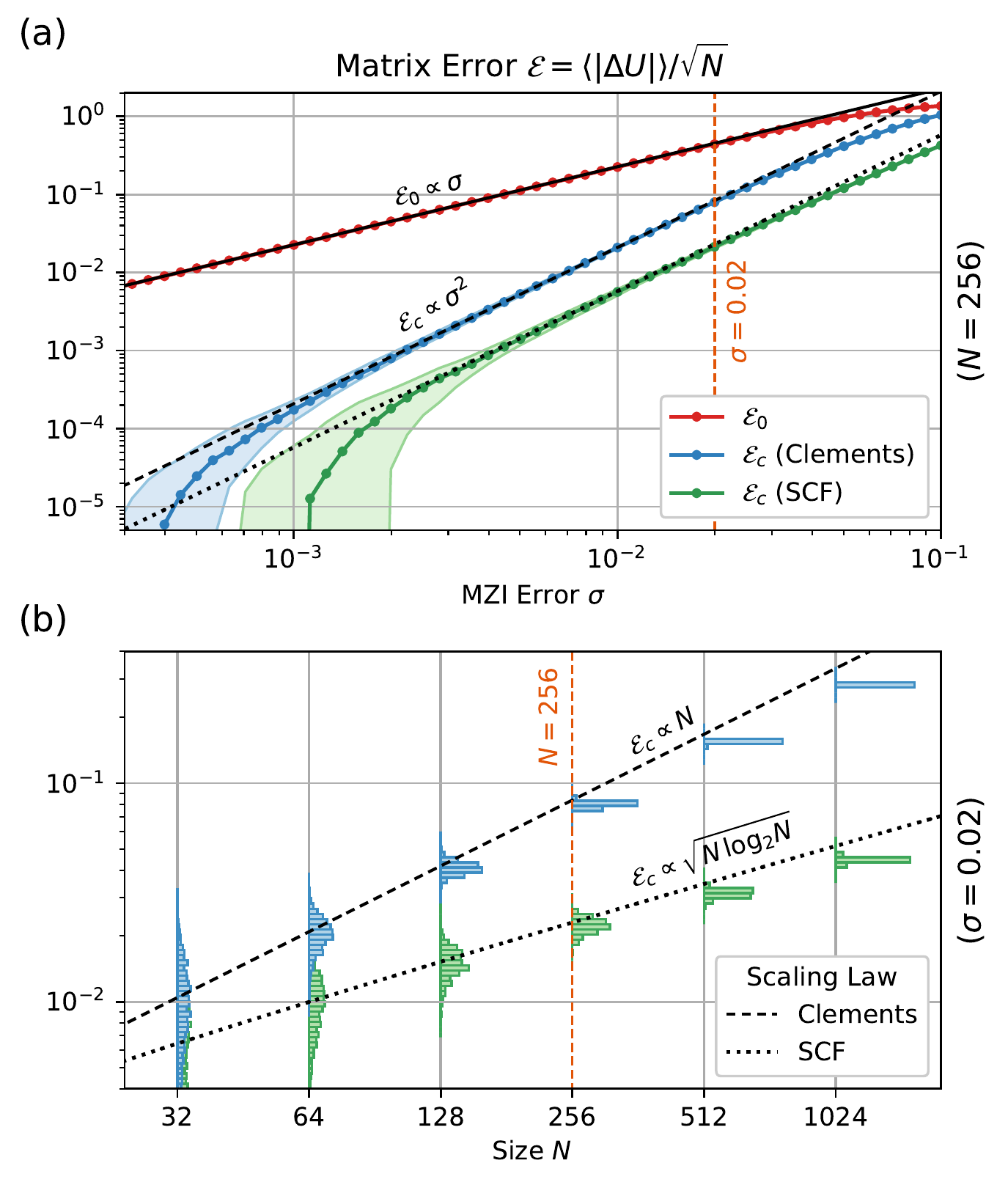}
    \caption{\textbf{(a)} Scaling of matrix error with process variation $\sigma$. \textbf{(b)} Scaling of matrix error with mesh size, showing the advantage of the SCF mesh over the \texttt{CLEMENTS} mesh for larger mesh sizes.}
    \label{fig_3}
\end{figure}

We performed numerical experiments on meshes up to size $N = 1024$ to validate the above expressions (Fig.~\ref{fig_3}). The observed uncorrected error $\mathcal{E}_0$ of both meshes (Fig.~\ref{fig_3}(a)) scales as $\sigma$ while error-correction on both meshes improves this scaling to $\sigma^{2}$. For meshes of size $256 \times 256$, error correction shows over an order of magnitude improvement in matrix error Frobenius norm over the uncorrected case when $\sigma \leq 0.01$, with the Sine-Cosine Fractal mesh performing better than the \texttt{CLEMENTS} mesh. Fig.~\ref{fig_3}(b) illustrates the distribution of post-correction error Frobenius norm as a function of mesh size $N$ for a fixed $\sigma = 0.02$ (which is a typical value for directional couplers under wafer-scale process variation). While the factor $\sqrt{\frac{N}{\mathrm{log}_{2}(N)}}$ is modest for small meshes ($N < 64$), it clearly opens up a significant accuracy gap in Fig.~\ref{fig_3}(b) in the large scale regime ($N > 1024$). 

\subsection{Fraction of Haar-random matrices that are exactly realizable in the presence of MZI errors}

The fraction of the unitary group $U(N)$ that can be realized by an imperfect mesh is equal to the probability that, under the Haar measure~\cite{MZI_rank}, all target splitting angles $\theta$ are realizable.  For convenience, we derive a quantity called the \textit{coverage}, cov(N)~\cite{self_con_1}, from this probability:
\begin{align}
        \text{cov}(N) &=\prod_n\left(1-P_n(\theta\!<\!\theta_{\rm min})-P_n(\theta\!>\!\theta_{\rm max})\right) \nonumber \\
    	& \approx \exp\Bigl(-\sum_n{\bigl[P_n(\theta\!<\!\theta_{\rm min}) + P_n(\theta\!>\!\theta_{\rm max})\bigr]}\Bigr)
    \label{eq:cov}
\end{align}
For a \texttt{CLEMENTS} mesh, Ref.~\cite{self_con_1} proves that $\text{cov}^{\text{(clem)}}(N) = \text{exp}\left( -\frac{N^{3}\sigma^{2}}{3} \right)$. Like the matrix error in the previous subsection, the coverage of an SCF mesh is computed using Taylor series expansions. The result is:
\begin{equation}
    \text{cov}^{\text{(scf)}}(N) = \exp\Bigl(-\frac{8N^2\log_2(N)}{\pi^2} \sigma^2\Bigr) \label{eq:covft}
\end{equation}
which is greater than $\text{cov}^{\text{(clem)}}(N)$ for the same $\sigma$ for all but very small $N$. 

\section{\label{sec:4}Use in Optical Neural Networks}

In this section, we study the performance of an optical neural network (ONN) built from Sine-Cosine Fractal meshes. We also propose and evaluate a pruning scheme for these meshes that allows areal footprint reduction while maintaining test performance. The neural network configuration is similar to those studied in Refs.~\cite{NN_arch_1, NN_arch_2, NN_arch_3}\textemdash each neural net layer is implemented by an SCF mesh connected to an electro-optic nonlinearity~\cite{MZI_NL} (Fig.~\ref{fig_5}(a)). All our networks had two layers. Our simulations used the \texttt{meshes}~\cite{meshes_ryan} package, and results are presented for the MNIST~\cite{MNIST} image classification task. The preprocessing of the images involved low-pass filtering and was identical to the procedure adopted in Refs.~\cite{NN_arch_3,hec_1}. The standard Cross Entropy loss and the Adam optimizer were used for training.

ONNs trained with SCF meshes achieved accuracies that matched the \texttt{CLEMENTS} mesh~\cite{hec_1, self_con_1} of $\sim$ 95\%-96\% for small meshes ($N = 64$) and $\sim$ 97\% for larger meshes ($N = 256$) . Next, we simulated the effect of MZI errors on the trained SCF mesh neural net; Fig.~\ref{fig_5}(b) shows the median classification accuracy of 10 independently trained networks as a function of splitter errors. SCF networks of size $N = 64$ yielded $\sim$ 95\% test accuracy while those of size $N = 256$ reached $\sim$ 97\%. The presence of MZI errors rapidly degrades the performance of the network, with accuracy dropping to below 90\% with splitter variation as low as $\sim$ 2\%. The use of hardware error correction, however, extends this cutoff to greater than 12\% even for bigger meshes, which is well above present-day process error~\cite{process_err} and larger than the corresponding cutoff for \texttt{CLEMENTS} meshes (which is 6\%~\cite{hec_1}).

\begin{figure}[t]
    \centering
    \includegraphics[width = \columnwidth]{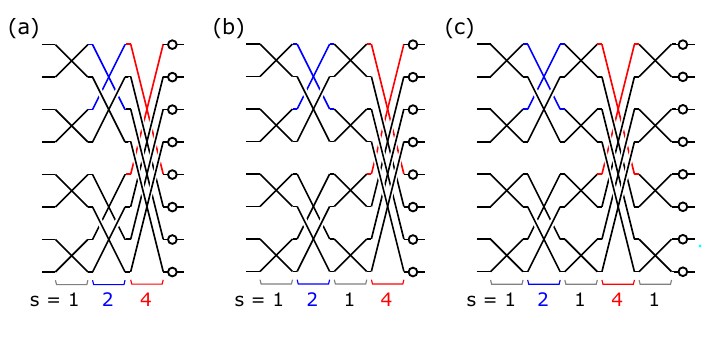}
    \caption{The SCF mesh at different stages of pruning: \textbf{(a)} A maximally pruned mesh, which is the standard FFT mesh, $\alpha = 1.0$. \textbf{(b)} $\alpha = 1.5$ \textbf{(c)} $\alpha = 1.75$ }
    \label{fig_4}
\end{figure}

\subsection*{Weight pruning}

The number of columns of MZIs with stride $s$ is $\frac{N}{2s}$ in the Sine-Cosine Fractal mesh while the standard FFT mesh contains a single column of each stride. We introduce a pruning scheme that interpolates between these extremes by introducing a \textit{fractal dimension} $\alpha \in [1, 2]$\textemdash in a partially pruned mesh, the number of columns of stride $s$ is $\left( \frac{N}{2s} \right)^{\alpha - 1}$. Setting $\alpha = 1$ and $\alpha = 2$ yields the FFT and SCF meshes respectively. Controlling $\alpha$ allows us to tune the number of degrees of freedom and reduce the areal footprint of the device while ensuring full connectivity. Fig.~\ref{fig_4} illustrates partially pruned $8 \times 8$ SCF meshes for different values of $\alpha$. The depth of the pruned network (approximated to leading order in $N$) scales as $\frac{N^{\alpha - 1}}{2^{\alpha - 1} - 1}$ while the number of MZIs is approximately $\frac{N^{\alpha}}{2\left(2^{\alpha} - 1\right)}$. Care has to be taken during pruning to ensure that no two consecutive columns have the same stride\textemdash any such columns would collapse into a single column.

\begin{figure}
    \centering
    \includegraphics[width = \columnwidth]{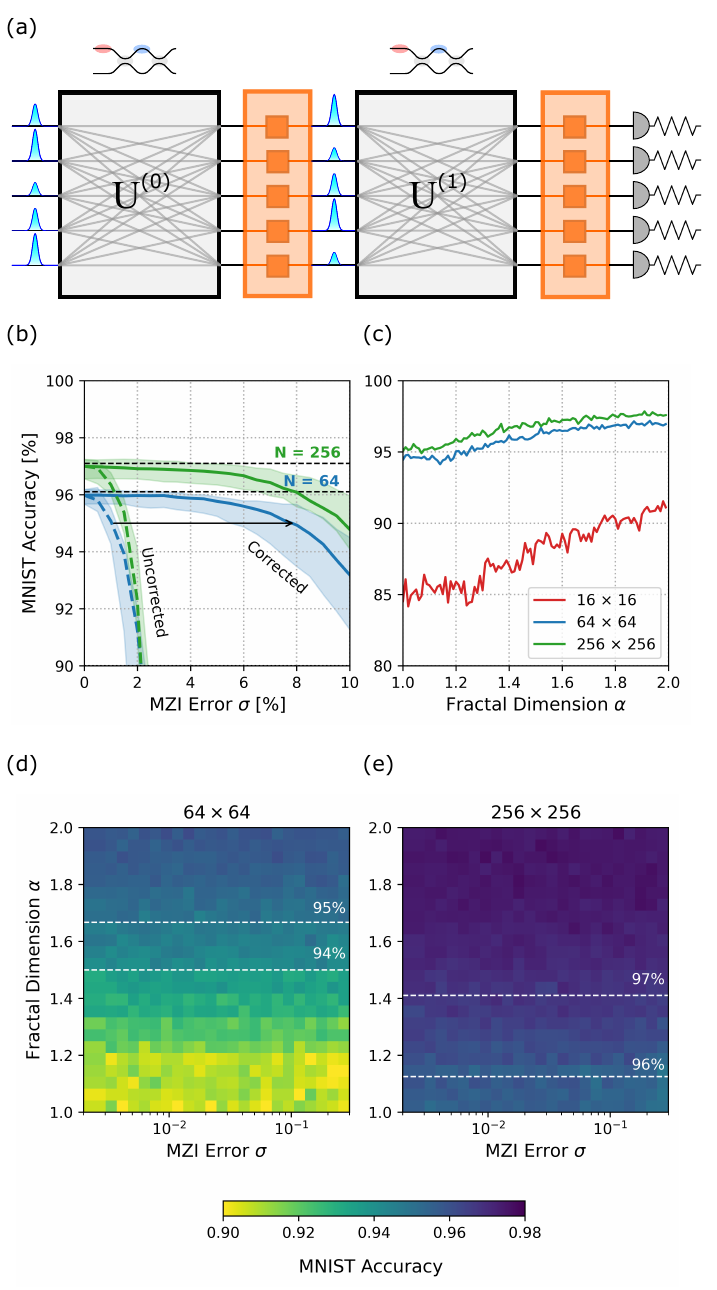}
    \caption{\textbf{(a)} 2-layer deep neural network architecture for MNIST classification where the blocks $U^{(0, 1)}$ represent unitary transforms by the Sine-Cosine Fractal mesh. \textbf{(b)} Simulated classification accuracy as a function of mesh size ($N =$ 64 and 256) and MZI error $\sigma$. \textbf{(c)} Classification accuracy as a function of degree of pruning $\alpha \in [1, 2]$ and mesh size ($N = $ 16 and 64 and 256). \textbf{(d)}, \textbf{(e)} Classification accuracy as a function of the degree of pruning $\alpha$ and MZI error, trained on maximally faulty mesh of size $N =$ 64 and 256.  }
    \label{fig_5}
\end{figure}

Fig.~\ref{fig_5}(c) illustrates the results of training networks with pruned meshes of ideal MZIs for different $\alpha$. Increasing $\alpha$ (decreasing the amount of pruning, or increasing the size of the network) increases the classification accuracy as one would expect. Interestingly, a maximally pruned (that is, a standard FFT) $64 \times 64$ 2-layer ONN already achieves 95\% accuracy, which is commensurate with the performance of present-day DNN accelerators~\cite{Opt_DNN_Ryan}.

\subsection*{Lower bounds on SCF mesh performance}

We performed error-aware `maximally faulty mesh' training \cite{Sri_paper} to obtain non-trivial (empirical) lower bounds on the performance of SCF mesh ONNs with MZI errors as large as $30\%$. Ref.~\cite{Sri_paper} shows that matrices obtained by training more faulty networks can be exactly transferred to less faulty ones. Maximally faulty 2-layer SCF mesh ONNs of sizes 64 and 256 were trained for several values of MZI error level $\sigma$ and fractal dimension $\alpha$; the resultant test accuracies are depicted in Fig.~\ref{fig_5}(d),(e). As expected, the 256 sized ONNs perform better than the 64 sized ones. However, it is particularly striking in the size 256 case that one can prune the network up to $\alpha\sim1.1$ and still lose only 1\% in classification accuracy. The size 64 case allows less freedom in pruning because it had a lower number of parameters to begin with. The performance also seems to be a nearly constant function of the MZI error. This investigation indicates that one can aggressively prune very faulty ONNs but still achieve excellent performance.   

\section{\label{sec:5}Conclusion}

This work presented a novel architecture for multiport interferometers based on the Sine-Cosine decomposition of unitary matrices. The proposed scheme is self-similar, and therefore modular. As a result, this architecture is ideal to construct multi-chiplet modules for large-scale devices that are typically limited by device yield. We showed that SCF meshes show improved scaling under error correction when compared to conventional multiport interferometers. Finally, this design allows for systematic re-wiring of MZI layers, which is an efficient way of reducing the areal footprint of the mesh while maintaining full connectivity. 

The proposed design has multiple advantages over the traditional architectures for multiport interferometers. Due to the uniform distribution of coupling angles (which is in contrast to conventional mesh architectures where crossing angles are clustered around the cross state), error correction techniques are far more effective in the case of SCF meshes. Truncated meshes used in ONNs can be trained to perform on par with the performances of present-day DNN accelerators. While the benefits of modularity and stronger robustness to error are small for present-day mesh size, the scaling gap of $O(N)$ vs. $O(\sqrt{N \log_{2}(N)})$ significantly impacts the performance for larger meshes. This reduced scaling with $N$ implies that SCF meshes are more senstive to improvements in foundry process. Reducing $\sigma$ by $2\times$ corresponds to a $4\times$ increase in the maximum mesh size for \textsc{Reck} and \textsc{Clements} owing to the $O(N\sigma^2)$ scaling; for the butterfly fractal, the corresponding increase would be $16\times$.

A potential drawback of this scheme is the presence of multiple large stride crossings with non-zero crosstalk. The loss/crosstalk introduced by these crossings can be minimized by the following methods that are enabled by the modularity of the SCF architecture:
\begin{itemize}
    \item \textit{Incomplete Decomposition}: To minimize the number of crossings, the block decomposition of the mesh can be terminated such that the smallest block size $N_{\mathrm{blk}} > 2$. Each block will then take the form of a standard \texttt{CLEMENTS} geometry with no intra-block crossings. 
    
    \item \textit{Out-of-plane Crossings}: Inter-block crossings ($s > 1$ among arbitrary block sizes) can be implemented using waveguide escalators and out-of-plane crossings. These crossings have been shown to have much lower crosstalk than in-plane crossings~\cite{3d_cross_1, 3d_cross_2}. In the case of a multi-chip module, each chiplet will be connected by a set of waveguide crossings with a large stride. These crossings can be fabricated using present-day lithography and laser-writing techniques in either polymer~\cite{3d_cross_polymer_1, 3d_cross_polymer_2} or glass~\cite{3d_cross_glass_1, 3d_cross_glass_2}, and could utilize ``hockey-stick'' escalator couplers to reduce the alignment tolerances for each chiplet \cite{HockeyStick}. 
\end{itemize}
This suggests that there exists an optimal block or chiplet size $N_{\mathrm{blk}}$ for multi-chip modules, which will have to be determined by the trade-off between intra-block errors (that favors small $N_{\mathrm{blk}}$) and inter-block losses (that favors large $N_{\mathrm{blk}}$).

For practical use as energy efficient deep learning accelerators, nanophotonic circuits will need to be scaled to reach the sub-fJ/MAC energy target. Present-day foundry processes face immense difficulties in scaling conventional multiport interferometers to the large size required to achieve this energy efficiency target. Furthermore, these meshes suffer unacceptably high errors $\mathcal{E}_c^{\mathrm{(clem)}} \gtrsim 0.2$ even in the presence of error correction. This brings us to the regime below the recommended 4--8 bits of precision that is usually targeted for DNN accelerators~\cite{DNN_acc_1, DNN_acc_2, DNN_acc_3}. The modularity of the SCF mesh, and its improved tolerance to error, open up the regime of large-scale programmable photonics, making the SCF mesh a frontrunning candidate architecture for future systems.  

\section*{Acknowledgements}

J.R.B.\ would like to acknowledge support from NTT Research, Inc.\ for the duration of this project. R.H., S.B., and D.E.\ would like to thank Prof.\ David A.B.\ Miller and Dr.\ Sunil Pai for helpful discussions. The simulations presented in this paper were performed on the DeepThought2 supercomputing cluster. The authors acknowledge the University of Maryland supercomputing resources (\url{http://hpcc.umd.edu}) made available for conducting the research reported in this paper.

\bibliography{main}% Produces the bibliography via BibTeX.

%apsrev4-2.bst 2019-01-14 (MD) hand-edited version of apsrev4-1.bst
%Control: key (0)
%Control: author (8) initials jnrlst
%Control: editor formatted (1) identically to author
%Control: production of article title (0) allowed
%Control: page (0) single
%Control: year (1) truncated
%Control: production of eprint (0) enabled
\begin{thebibliography}{56}%
\makeatletter
\providecommand \@ifxundefined [1]{%
 \@ifx{#1\undefined}
}%
\providecommand \@ifnum [1]{%
 \ifnum #1\expandafter \@firstoftwo
 \else \expandafter \@secondoftwo
 \fi
}%
\providecommand \@ifx [1]{%
 \ifx #1\expandafter \@firstoftwo
 \else \expandafter \@secondoftwo
 \fi
}%
\providecommand \natexlab [1]{#1}%
\providecommand \enquote  [1]{``#1''}%
\providecommand \bibnamefont  [1]{#1}%
\providecommand \bibfnamefont [1]{#1}%
\providecommand \citenamefont [1]{#1}%
\providecommand \href@noop [0]{\@secondoftwo}%
\providecommand \href [0]{\begingroup \@sanitize@url \@href}%
\providecommand \@href[1]{\@@startlink{#1}\@@href}%
\providecommand \@@href[1]{\endgroup#1\@@endlink}%
\providecommand \@sanitize@url [0]{\catcode `\\12\catcode `\$12\catcode
  `\&12\catcode `\#12\catcode `\^12\catcode `\_12\catcode `\%12\relax}%
\providecommand \@@startlink[1]{}%
\providecommand \@@endlink[0]{}%
\providecommand \url  [0]{\begingroup\@sanitize@url \@url }%
\providecommand \@url [1]{\endgroup\@href {#1}{\urlprefix }}%
\providecommand \urlprefix  [0]{URL }%
\providecommand \Eprint [0]{\href }%
\providecommand \doibase [0]{https://doi.org/}%
\providecommand \selectlanguage [0]{\@gobble}%
\providecommand \bibinfo  [0]{\@secondoftwo}%
\providecommand \bibfield  [0]{\@secondoftwo}%
\providecommand \translation [1]{[#1]}%
\providecommand \BibitemOpen [0]{}%
\providecommand \bibitemStop [0]{}%
\providecommand \bibitemNoStop [0]{.\EOS\space}%
\providecommand \EOS [0]{\spacefactor3000\relax}%
\providecommand \BibitemShut  [1]{\csname bibitem#1\endcsname}%
\let\auto@bib@innerbib\@empty
%</preamble>
\bibitem [{\citenamefont {Annoni}\ \emph {et~al.}(2017)\citenamefont {Annoni},
  \citenamefont {Guglielmi}, \citenamefont {Carminati}, \citenamefont
  {Ferrari}, \citenamefont {Sampietro}, \citenamefont {Miller}, \citenamefont
  {Melloni},\ and\ \citenamefont {Morichetti}}]{annoni2017unscrambling}%
  \BibitemOpen
  \bibfield  {author} {\bibinfo {author} {\bibfnamefont {A.}~\bibnamefont
  {Annoni}}, \bibinfo {author} {\bibfnamefont {E.}~\bibnamefont {Guglielmi}},
  \bibinfo {author} {\bibfnamefont {M.}~\bibnamefont {Carminati}}, \bibinfo
  {author} {\bibfnamefont {G.}~\bibnamefont {Ferrari}}, \bibinfo {author}
  {\bibfnamefont {M.}~\bibnamefont {Sampietro}}, \bibinfo {author}
  {\bibfnamefont {D.~A.}\ \bibnamefont {Miller}}, \bibinfo {author}
  {\bibfnamefont {A.}~\bibnamefont {Melloni}},\ and\ \bibinfo {author}
  {\bibfnamefont {F.}~\bibnamefont {Morichetti}},\ }\bibfield  {title}
  {\bibinfo {title} {{Unscrambling light—automatically undoing strong mixing
  between modes}},\ }\href {https://doi.org/10.1038/lsa.2017.110} {\bibfield
  {journal} {\bibinfo  {journal} {Light: Science \& Applications}\ }\textbf
  {\bibinfo {volume} {6}},\ \bibinfo {pages} {e17110} (\bibinfo {year}
  {2017})}\BibitemShut {NoStop}%
\bibitem [{\citenamefont {Ribeiro}\ \emph {et~al.}(2016)\citenamefont
  {Ribeiro}, \citenamefont {Ruocco}, \citenamefont {Vanacker},\ and\
  \citenamefont {Bogaerts}}]{ribeiro2016demonstration}%
  \BibitemOpen
  \bibfield  {author} {\bibinfo {author} {\bibfnamefont {A.}~\bibnamefont
  {Ribeiro}}, \bibinfo {author} {\bibfnamefont {A.}~\bibnamefont {Ruocco}},
  \bibinfo {author} {\bibfnamefont {L.}~\bibnamefont {Vanacker}},\ and\
  \bibinfo {author} {\bibfnamefont {W.}~\bibnamefont {Bogaerts}},\ }\bibfield
  {title} {\bibinfo {title} {{Demonstration of a 4$\times$ 4-port universal
  linear circuit}},\ }\href {https://doi.org/10.1364/OPTICA.3.001348}
  {\bibfield  {journal} {\bibinfo  {journal} {Optica}\ }\textbf {\bibinfo
  {volume} {3}},\ \bibinfo {pages} {1348} (\bibinfo {year} {2016})}\BibitemShut
  {NoStop}%
\bibitem [{\citenamefont {Milanizadeh}\ \emph {et~al.}(2019)\citenamefont
  {Milanizadeh}, \citenamefont {Borga}, \citenamefont {Morichetti},
  \citenamefont {Miller},\ and\ \citenamefont
  {Melloni}}]{milanizadeh2019manipulating}%
  \BibitemOpen
  \bibfield  {author} {\bibinfo {author} {\bibfnamefont {M.}~\bibnamefont
  {Milanizadeh}}, \bibinfo {author} {\bibfnamefont {P.}~\bibnamefont {Borga}},
  \bibinfo {author} {\bibfnamefont {F.}~\bibnamefont {Morichetti}}, \bibinfo
  {author} {\bibfnamefont {D.}~\bibnamefont {Miller}},\ and\ \bibinfo {author}
  {\bibfnamefont {A.}~\bibnamefont {Melloni}},\ }\bibfield  {title} {\bibinfo
  {title} {{Manipulating free-space optical beams with a silicon photonic
  mesh}},\ }in\ \href {https://doi.org/10.1109/PHOSST.2019.8795053} {\emph
  {\bibinfo {booktitle} {2019 IEEE Photonics Society Summer Topical Meeting
  Series (SUM)}}}\ (\bibinfo {organization} {IEEE},\ \bibinfo {year} {2019})\
  pp.\ \bibinfo {pages} {1--2}\BibitemShut {NoStop}%
\bibitem [{\citenamefont {Zhuang}\ \emph {et~al.}(2015)\citenamefont {Zhuang},
  \citenamefont {Roeloffzen}, \citenamefont {Hoekman}, \citenamefont {Boller},\
  and\ \citenamefont {Lowery}}]{zhuang2015programmable}%
  \BibitemOpen
  \bibfield  {author} {\bibinfo {author} {\bibfnamefont {L.}~\bibnamefont
  {Zhuang}}, \bibinfo {author} {\bibfnamefont {C.~G.}\ \bibnamefont
  {Roeloffzen}}, \bibinfo {author} {\bibfnamefont {M.}~\bibnamefont {Hoekman}},
  \bibinfo {author} {\bibfnamefont {K.-J.}\ \bibnamefont {Boller}},\ and\
  \bibinfo {author} {\bibfnamefont {A.~J.}\ \bibnamefont {Lowery}},\ }\bibfield
   {title} {\bibinfo {title} {{Programmable photonic signal processor chip for
  radiofrequency applications}},\ }\href
  {https://doi.org/10.1364/OPTICA.2.000854} {\bibfield  {journal} {\bibinfo
  {journal} {Optica}\ }\textbf {\bibinfo {volume} {2}},\ \bibinfo {pages} {854}
  (\bibinfo {year} {2015})}\BibitemShut {NoStop}%
\bibitem [{\citenamefont {Notaros}\ \emph {et~al.}(2017)\citenamefont
  {Notaros}, \citenamefont {Mower}, \citenamefont {Heuck}, \citenamefont
  {Lupo}, \citenamefont {Harris}, \citenamefont {Steinbrecher}, \citenamefont
  {Bunandar}, \citenamefont {Baehr-Jones}, \citenamefont {Hochberg},
  \citenamefont {Lloyd} \emph {et~al.}}]{notaros2017programmable}%
  \BibitemOpen
  \bibfield  {author} {\bibinfo {author} {\bibfnamefont {J.}~\bibnamefont
  {Notaros}}, \bibinfo {author} {\bibfnamefont {J.}~\bibnamefont {Mower}},
  \bibinfo {author} {\bibfnamefont {M.}~\bibnamefont {Heuck}}, \bibinfo
  {author} {\bibfnamefont {C.}~\bibnamefont {Lupo}}, \bibinfo {author}
  {\bibfnamefont {N.~C.}\ \bibnamefont {Harris}}, \bibinfo {author}
  {\bibfnamefont {G.~R.}\ \bibnamefont {Steinbrecher}}, \bibinfo {author}
  {\bibfnamefont {D.}~\bibnamefont {Bunandar}}, \bibinfo {author}
  {\bibfnamefont {T.}~\bibnamefont {Baehr-Jones}}, \bibinfo {author}
  {\bibfnamefont {M.}~\bibnamefont {Hochberg}}, \bibinfo {author}
  {\bibfnamefont {S.}~\bibnamefont {Lloyd}}, \emph {et~al.},\ }\bibfield
  {title} {\bibinfo {title} {{Programmable dispersion on a photonic integrated
  circuit for classical and quantum applications}},\ }\href
  {https://doi.org/10.1364/OE.25.021275} {\bibfield  {journal} {\bibinfo
  {journal} {Optics Express}\ }\textbf {\bibinfo {volume} {25}},\ \bibinfo
  {pages} {21275} (\bibinfo {year} {2017})}\BibitemShut {NoStop}%
\bibitem [{\citenamefont {Shen}\ \emph {et~al.}(2017)\citenamefont {Shen},
  \citenamefont {Harris}, \citenamefont {Skirlo}, \citenamefont {Prabhu},
  \citenamefont {Baehr-Jones}, \citenamefont {Hochberg}, \citenamefont {Sun},
  \citenamefont {Zhao}, \citenamefont {Larochelle}, \citenamefont {Englund}
  \emph {et~al.}}]{NN_arch_1}%
  \BibitemOpen
  \bibfield  {author} {\bibinfo {author} {\bibfnamefont {Y.}~\bibnamefont
  {Shen}}, \bibinfo {author} {\bibfnamefont {N.~C.}\ \bibnamefont {Harris}},
  \bibinfo {author} {\bibfnamefont {S.}~\bibnamefont {Skirlo}}, \bibinfo
  {author} {\bibfnamefont {M.}~\bibnamefont {Prabhu}}, \bibinfo {author}
  {\bibfnamefont {T.}~\bibnamefont {Baehr-Jones}}, \bibinfo {author}
  {\bibfnamefont {M.}~\bibnamefont {Hochberg}}, \bibinfo {author}
  {\bibfnamefont {X.}~\bibnamefont {Sun}}, \bibinfo {author} {\bibfnamefont
  {S.}~\bibnamefont {Zhao}}, \bibinfo {author} {\bibfnamefont {H.}~\bibnamefont
  {Larochelle}}, \bibinfo {author} {\bibfnamefont {D.}~\bibnamefont {Englund}},
  \emph {et~al.},\ }\bibfield  {title} {\bibinfo {title} {{Deep learning with
  coherent nanophotonic circuits}},\ }\href
  {https://doi.org/10.1038/nphoton.2017.93} {\bibfield  {journal} {\bibinfo
  {journal} {Nature Photonics}\ }\textbf {\bibinfo {volume} {11}},\ \bibinfo
  {pages} {441} (\bibinfo {year} {2017})}\BibitemShut {NoStop}%
\bibitem [{\citenamefont {Basani}\ \emph {et~al.}(2022)\citenamefont {Basani},
  \citenamefont {Heuck}, \citenamefont {Englund},\ and\ \citenamefont
  {Krastanov}}]{opt_ml}%
  \BibitemOpen
  \bibfield  {author} {\bibinfo {author} {\bibfnamefont {J.~R.}\ \bibnamefont
  {Basani}}, \bibinfo {author} {\bibfnamefont {M.}~\bibnamefont {Heuck}},
  \bibinfo {author} {\bibfnamefont {D.~R.}\ \bibnamefont {Englund}},\ and\
  \bibinfo {author} {\bibfnamefont {S.}~\bibnamefont {Krastanov}},\ }\bibfield
  {title} {\bibinfo {title} {All-photonic artificial neural network processor
  via non-linear optics},\ }\bibfield  {journal} {\bibinfo  {journal} {arXiv
  preprint arXiv:2205.08608}\ }\href {https://doi.org/arXiv:2205.08608}
  {arXiv:2205.08608} (\bibinfo {year} {2022})\BibitemShut {NoStop}%
\bibitem [{\citenamefont {Prabhu}\ \emph {et~al.}(2020)\citenamefont {Prabhu},
  \citenamefont {Roques-Carmes}, \citenamefont {Shen}, \citenamefont {Harris},
  \citenamefont {Jing}, \citenamefont {Carolan}, \citenamefont {Hamerly},
  \citenamefont {Baehr-Jones}, \citenamefont {Hochberg}, \citenamefont
  {{\v{C}}eperi{\'c}} \emph {et~al.}}]{prabhu2020ising}%
  \BibitemOpen
  \bibfield  {author} {\bibinfo {author} {\bibfnamefont {M.}~\bibnamefont
  {Prabhu}}, \bibinfo {author} {\bibfnamefont {C.}~\bibnamefont
  {Roques-Carmes}}, \bibinfo {author} {\bibfnamefont {Y.}~\bibnamefont {Shen}},
  \bibinfo {author} {\bibfnamefont {N.}~\bibnamefont {Harris}}, \bibinfo
  {author} {\bibfnamefont {L.}~\bibnamefont {Jing}}, \bibinfo {author}
  {\bibfnamefont {J.}~\bibnamefont {Carolan}}, \bibinfo {author} {\bibfnamefont
  {R.}~\bibnamefont {Hamerly}}, \bibinfo {author} {\bibfnamefont
  {T.}~\bibnamefont {Baehr-Jones}}, \bibinfo {author} {\bibfnamefont
  {M.}~\bibnamefont {Hochberg}}, \bibinfo {author} {\bibfnamefont
  {V.}~\bibnamefont {{\v{C}}eperi{\'c}}}, \emph {et~al.},\ }\bibfield  {title}
  {\bibinfo {title} {Accelerating recurrent ising machines in photonic
  integrated circuits},\ }\href {https://doi.org/10.1364/OPTICA.386613}
  {\bibfield  {journal} {\bibinfo  {journal} {Optica}\ }\textbf {\bibinfo
  {volume} {7}},\ \bibinfo {pages} {551} (\bibinfo {year} {2020})}\BibitemShut
  {NoStop}%
\bibitem [{\citenamefont {Harris}\ \emph {et~al.}(2017)\citenamefont {Harris},
  \citenamefont {Steinbrecher}, \citenamefont {Prabhu}, \citenamefont {Lahini},
  \citenamefont {Mower}, \citenamefont {Bunandar}, \citenamefont {Chen},
  \citenamefont {Wong}, \citenamefont {Baehr-Jones}, \citenamefont {Hochberg}
  \emph {et~al.}}]{harris2017quantum}%
  \BibitemOpen
  \bibfield  {author} {\bibinfo {author} {\bibfnamefont {N.~C.}\ \bibnamefont
  {Harris}}, \bibinfo {author} {\bibfnamefont {G.~R.}\ \bibnamefont
  {Steinbrecher}}, \bibinfo {author} {\bibfnamefont {M.}~\bibnamefont
  {Prabhu}}, \bibinfo {author} {\bibfnamefont {Y.}~\bibnamefont {Lahini}},
  \bibinfo {author} {\bibfnamefont {J.}~\bibnamefont {Mower}}, \bibinfo
  {author} {\bibfnamefont {D.}~\bibnamefont {Bunandar}}, \bibinfo {author}
  {\bibfnamefont {C.}~\bibnamefont {Chen}}, \bibinfo {author} {\bibfnamefont
  {F.~N.}\ \bibnamefont {Wong}}, \bibinfo {author} {\bibfnamefont
  {T.}~\bibnamefont {Baehr-Jones}}, \bibinfo {author} {\bibfnamefont
  {M.}~\bibnamefont {Hochberg}}, \emph {et~al.},\ }\bibfield  {title} {\bibinfo
  {title} {{Quantum transport simulations in a programmable nanophotonic
  processor}},\ }\href {https://doi.org/10.1038/nphoton.2017.95} {\bibfield
  {journal} {\bibinfo  {journal} {Nature Photonics}\ }\textbf {\bibinfo
  {volume} {11}},\ \bibinfo {pages} {447} (\bibinfo {year} {2017})}\BibitemShut
  {NoStop}%
\bibitem [{\citenamefont {Wang}\ \emph {et~al.}(2018)\citenamefont {Wang},
  \citenamefont {Paesani}, \citenamefont {Ding}, \citenamefont {Santagati},
  \citenamefont {Skrzypczyk}, \citenamefont {Salavrakos}, \citenamefont {Tura},
  \citenamefont {Augusiak}, \citenamefont {Man{\v{c}}inska}, \citenamefont
  {Bacco} \emph {et~al.}}]{wang2018multidimensional}%
  \BibitemOpen
  \bibfield  {author} {\bibinfo {author} {\bibfnamefont {J.}~\bibnamefont
  {Wang}}, \bibinfo {author} {\bibfnamefont {S.}~\bibnamefont {Paesani}},
  \bibinfo {author} {\bibfnamefont {Y.}~\bibnamefont {Ding}}, \bibinfo {author}
  {\bibfnamefont {R.}~\bibnamefont {Santagati}}, \bibinfo {author}
  {\bibfnamefont {P.}~\bibnamefont {Skrzypczyk}}, \bibinfo {author}
  {\bibfnamefont {A.}~\bibnamefont {Salavrakos}}, \bibinfo {author}
  {\bibfnamefont {J.}~\bibnamefont {Tura}}, \bibinfo {author} {\bibfnamefont
  {R.}~\bibnamefont {Augusiak}}, \bibinfo {author} {\bibfnamefont
  {L.}~\bibnamefont {Man{\v{c}}inska}}, \bibinfo {author} {\bibfnamefont
  {D.}~\bibnamefont {Bacco}}, \emph {et~al.},\ }\bibfield  {title} {\bibinfo
  {title} {{Multidimensional quantum entanglement with large-scale integrated
  optics}},\ }\href {https://doi.org/10.1126/science.aar7053} {\bibfield
  {journal} {\bibinfo  {journal} {Science}\ }\textbf {\bibinfo {volume}
  {360}},\ \bibinfo {pages} {285} (\bibinfo {year} {2018})}\BibitemShut
  {NoStop}%
\bibitem [{\citenamefont {Qiang}\ \emph {et~al.}(2018)\citenamefont {Qiang},
  \citenamefont {Zhou}, \citenamefont {Wang}, \citenamefont {Wilkes},
  \citenamefont {Loke}, \citenamefont {O’Gara}, \citenamefont {Kling},
  \citenamefont {Marshall}, \citenamefont {Santagati}, \citenamefont {Ralph}
  \emph {et~al.}}]{qiang2018large}%
  \BibitemOpen
  \bibfield  {author} {\bibinfo {author} {\bibfnamefont {X.}~\bibnamefont
  {Qiang}}, \bibinfo {author} {\bibfnamefont {X.}~\bibnamefont {Zhou}},
  \bibinfo {author} {\bibfnamefont {J.}~\bibnamefont {Wang}}, \bibinfo {author}
  {\bibfnamefont {C.~M.}\ \bibnamefont {Wilkes}}, \bibinfo {author}
  {\bibfnamefont {T.}~\bibnamefont {Loke}}, \bibinfo {author} {\bibfnamefont
  {S.}~\bibnamefont {O’Gara}}, \bibinfo {author} {\bibfnamefont
  {L.}~\bibnamefont {Kling}}, \bibinfo {author} {\bibfnamefont {G.~D.}\
  \bibnamefont {Marshall}}, \bibinfo {author} {\bibfnamefont {R.}~\bibnamefont
  {Santagati}}, \bibinfo {author} {\bibfnamefont {T.~C.}\ \bibnamefont
  {Ralph}}, \emph {et~al.},\ }\bibfield  {title} {\bibinfo {title}
  {{Large-scale silicon quantum photonics implementing arbitrary two-qubit
  processing}},\ }\href {https://doi.org/10.1038/s41566-018-0236-y} {\bibfield
  {journal} {\bibinfo  {journal} {Nature Photonics}\ }\textbf {\bibinfo
  {volume} {12}},\ \bibinfo {pages} {534} (\bibinfo {year} {2018})}\BibitemShut
  {NoStop}%
\bibitem [{\citenamefont {Sparrow}\ \emph {et~al.}(2018)\citenamefont
  {Sparrow}, \citenamefont {Mart{\'\i}n-L{\'o}pez}, \citenamefont {Maraviglia},
  \citenamefont {Neville}, \citenamefont {Harrold}, \citenamefont {Carolan},
  \citenamefont {Joglekar}, \citenamefont {Hashimoto}, \citenamefont {Matsuda},
  \citenamefont {O’Brien} \emph {et~al.}}]{sparrow2018simulating}%
  \BibitemOpen
  \bibfield  {author} {\bibinfo {author} {\bibfnamefont {C.}~\bibnamefont
  {Sparrow}}, \bibinfo {author} {\bibfnamefont {E.}~\bibnamefont
  {Mart{\'\i}n-L{\'o}pez}}, \bibinfo {author} {\bibfnamefont {N.}~\bibnamefont
  {Maraviglia}}, \bibinfo {author} {\bibfnamefont {A.}~\bibnamefont {Neville}},
  \bibinfo {author} {\bibfnamefont {C.}~\bibnamefont {Harrold}}, \bibinfo
  {author} {\bibfnamefont {J.}~\bibnamefont {Carolan}}, \bibinfo {author}
  {\bibfnamefont {Y.~N.}\ \bibnamefont {Joglekar}}, \bibinfo {author}
  {\bibfnamefont {T.}~\bibnamefont {Hashimoto}}, \bibinfo {author}
  {\bibfnamefont {N.}~\bibnamefont {Matsuda}}, \bibinfo {author} {\bibfnamefont
  {J.~L.}\ \bibnamefont {O’Brien}}, \emph {et~al.},\ }\bibfield  {title}
  {\bibinfo {title} {{Simulating the vibrational quantum dynamics of molecules
  using photonics}},\ }\href {https://doi.org/10.1038/s41586-018-0152-9}
  {\bibfield  {journal} {\bibinfo  {journal} {Nature}\ }\textbf {\bibinfo
  {volume} {557}},\ \bibinfo {pages} {660} (\bibinfo {year}
  {2018})}\BibitemShut {NoStop}%
\bibitem [{\citenamefont {Carolan}\ \emph {et~al.}(2015)\citenamefont
  {Carolan}, \citenamefont {Harrold}, \citenamefont {Sparrow}, \citenamefont
  {Mart{\'\i}n-L{\'o}pez}, \citenamefont {Russell}, \citenamefont
  {Silverstone}, \citenamefont {Shadbolt}, \citenamefont {Matsuda},
  \citenamefont {Oguma}, \citenamefont {Itoh} \emph
  {et~al.}}]{carolan2015universal}%
  \BibitemOpen
  \bibfield  {author} {\bibinfo {author} {\bibfnamefont {J.}~\bibnamefont
  {Carolan}}, \bibinfo {author} {\bibfnamefont {C.}~\bibnamefont {Harrold}},
  \bibinfo {author} {\bibfnamefont {C.}~\bibnamefont {Sparrow}}, \bibinfo
  {author} {\bibfnamefont {E.}~\bibnamefont {Mart{\'\i}n-L{\'o}pez}}, \bibinfo
  {author} {\bibfnamefont {N.~J.}\ \bibnamefont {Russell}}, \bibinfo {author}
  {\bibfnamefont {J.~W.}\ \bibnamefont {Silverstone}}, \bibinfo {author}
  {\bibfnamefont {P.~J.}\ \bibnamefont {Shadbolt}}, \bibinfo {author}
  {\bibfnamefont {N.}~\bibnamefont {Matsuda}}, \bibinfo {author} {\bibfnamefont
  {M.}~\bibnamefont {Oguma}}, \bibinfo {author} {\bibfnamefont
  {M.}~\bibnamefont {Itoh}}, \emph {et~al.},\ }\bibfield  {title} {\bibinfo
  {title} {{Universal linear optics}},\ }\href
  {https://doi.org/10.1126/science.aab3642} {\bibfield  {journal} {\bibinfo
  {journal} {Science}\ }\textbf {\bibinfo {volume} {349}},\ \bibinfo {pages}
  {711} (\bibinfo {year} {2015})}\BibitemShut {NoStop}%
\bibitem [{\citenamefont {Fang}\ \emph {et~al.}(2019)\citenamefont {Fang},
  \citenamefont {Manipatruni}, \citenamefont {Wierzynski}, \citenamefont
  {Khosrowshahi},\ and\ \citenamefont {DeWeese}}]{NN_arch_2}%
  \BibitemOpen
  \bibfield  {author} {\bibinfo {author} {\bibfnamefont {M.~Y.-S.}\
  \bibnamefont {Fang}}, \bibinfo {author} {\bibfnamefont {S.}~\bibnamefont
  {Manipatruni}}, \bibinfo {author} {\bibfnamefont {C.}~\bibnamefont
  {Wierzynski}}, \bibinfo {author} {\bibfnamefont {A.}~\bibnamefont
  {Khosrowshahi}},\ and\ \bibinfo {author} {\bibfnamefont {M.~R.}\ \bibnamefont
  {DeWeese}},\ }\bibfield  {title} {\bibinfo {title} {{Design of optical neural
  networks with component imprecisions}},\ }\href
  {https://doi.org/10.1364/OE.27.014009} {\bibfield  {journal} {\bibinfo
  {journal} {Optics Express}\ }\textbf {\bibinfo {volume} {27}},\ \bibinfo
  {pages} {14009} (\bibinfo {year} {2019})}\BibitemShut {NoStop}%
\bibitem [{\citenamefont {Burgwal}\ \emph {et~al.}(2017)\citenamefont
  {Burgwal}, \citenamefont {Clements}, \citenamefont {Smith}, \citenamefont
  {Gates}, \citenamefont {Kolthammer}, \citenamefont {Renema},\ and\
  \citenamefont {Walmsley}}]{imperfect_mesh_1}%
  \BibitemOpen
  \bibfield  {author} {\bibinfo {author} {\bibfnamefont {R.}~\bibnamefont
  {Burgwal}}, \bibinfo {author} {\bibfnamefont {W.~R.}\ \bibnamefont
  {Clements}}, \bibinfo {author} {\bibfnamefont {D.~H.}\ \bibnamefont {Smith}},
  \bibinfo {author} {\bibfnamefont {J.~C.}\ \bibnamefont {Gates}}, \bibinfo
  {author} {\bibfnamefont {W.~S.}\ \bibnamefont {Kolthammer}}, \bibinfo
  {author} {\bibfnamefont {J.~J.}\ \bibnamefont {Renema}},\ and\ \bibinfo
  {author} {\bibfnamefont {I.~A.}\ \bibnamefont {Walmsley}},\ }\bibfield
  {title} {\bibinfo {title} {{Using an imperfect photonic network to implement
  random unitaries}},\ }\href {https://doi.org/10.1364/OE.25.028236} {\bibfield
   {journal} {\bibinfo  {journal} {Optics Express}\ }\textbf {\bibinfo {volume}
  {25}},\ \bibinfo {pages} {28236} (\bibinfo {year} {2017})}\BibitemShut
  {NoStop}%
\bibitem [{\citenamefont {Mower}\ \emph {et~al.}(2015)\citenamefont {Mower},
  \citenamefont {Harris}, \citenamefont {Steinbrecher}, \citenamefont
  {Lahini},\ and\ \citenamefont {Englund}}]{imperfect_mesh_2}%
  \BibitemOpen
  \bibfield  {author} {\bibinfo {author} {\bibfnamefont {J.}~\bibnamefont
  {Mower}}, \bibinfo {author} {\bibfnamefont {N.~C.}\ \bibnamefont {Harris}},
  \bibinfo {author} {\bibfnamefont {G.~R.}\ \bibnamefont {Steinbrecher}},
  \bibinfo {author} {\bibfnamefont {Y.}~\bibnamefont {Lahini}},\ and\ \bibinfo
  {author} {\bibfnamefont {D.}~\bibnamefont {Englund}},\ }\bibfield  {title}
  {\bibinfo {title} {{High-fidelity quantum state evolution in imperfect
  photonic integrated circuits}},\ }\href
  {https://doi.org/10.1103/PhysRevA.92.032322} {\bibfield  {journal} {\bibinfo
  {journal} {Physical Review A}\ }\textbf {\bibinfo {volume} {92}},\ \bibinfo
  {pages} {032322} (\bibinfo {year} {2015})}\BibitemShut {NoStop}%
\bibitem [{\citenamefont {L{\'o}pez}(2019)}]{lopez2019programmable}%
  \BibitemOpen
  \bibfield  {author} {\bibinfo {author} {\bibfnamefont {D.~P.}\ \bibnamefont
  {L{\'o}pez}},\ }\bibfield  {title} {\bibinfo {title} {{Programmable
  integrated silicon photonics waveguide meshes: optimized designs and control
  algorithms}},\ }\href {https://doi.org/10.1109/JSTQE.2019.2948048} {\bibfield
   {journal} {\bibinfo  {journal} {IEEE Journal of Selected Topics in Quantum
  Electronics}\ }\textbf {\bibinfo {volume} {26}},\ \bibinfo {pages} {1}
  (\bibinfo {year} {2019})}\BibitemShut {NoStop}%
\bibitem [{\citenamefont {L{\'o}pez}\ \emph {et~al.}(2020)\citenamefont
  {L{\'o}pez}, \citenamefont {P{\'e}rez}, \citenamefont {DasMahapatra},\ and\
  \citenamefont {Capmany}}]{lopez2020auto}%
  \BibitemOpen
  \bibfield  {author} {\bibinfo {author} {\bibfnamefont {A.}~\bibnamefont
  {L{\'o}pez}}, \bibinfo {author} {\bibfnamefont {D.}~\bibnamefont
  {P{\'e}rez}}, \bibinfo {author} {\bibfnamefont {P.}~\bibnamefont
  {DasMahapatra}},\ and\ \bibinfo {author} {\bibfnamefont {J.}~\bibnamefont
  {Capmany}},\ }\bibfield  {title} {\bibinfo {title} {{Auto-routing algorithm
  for field-programmable photonic gate arrays}},\ }\href
  {https://doi.org/10.1364/OE.382753} {\bibfield  {journal} {\bibinfo
  {journal} {Optics Express}\ }\textbf {\bibinfo {volume} {28}},\ \bibinfo
  {pages} {737} (\bibinfo {year} {2020})}\BibitemShut {NoStop}%
\bibitem [{\citenamefont {P{\'e}rez-L{\'o}pez}\ \emph
  {et~al.}(2020)\citenamefont {P{\'e}rez-L{\'o}pez}, \citenamefont {L{\'o}pez},
  \citenamefont {DasMahapatra},\ and\ \citenamefont
  {Capmany}}]{perez2020multipurpose}%
  \BibitemOpen
  \bibfield  {author} {\bibinfo {author} {\bibfnamefont {D.}~\bibnamefont
  {P{\'e}rez-L{\'o}pez}}, \bibinfo {author} {\bibfnamefont {A.}~\bibnamefont
  {L{\'o}pez}}, \bibinfo {author} {\bibfnamefont {P.}~\bibnamefont
  {DasMahapatra}},\ and\ \bibinfo {author} {\bibfnamefont {J.}~\bibnamefont
  {Capmany}},\ }\bibfield  {title} {\bibinfo {title} {{Multipurpose
  self-configuration of programmable photonic circuits}},\ }\href
  {https://doi.org/10.1038/s41467-020-19608-w} {\bibfield  {journal} {\bibinfo
  {journal} {Nature Communications}\ }\textbf {\bibinfo {volume} {11}},\
  \bibinfo {pages} {1} (\bibinfo {year} {2020})}\BibitemShut {NoStop}%
\bibitem [{\citenamefont {Pai}\ \emph {et~al.}(2019)\citenamefont {Pai},
  \citenamefont {Bartlett}, \citenamefont {Solgaard},\ and\ \citenamefont
  {Miller}}]{imperfect_mesh_3}%
  \BibitemOpen
  \bibfield  {author} {\bibinfo {author} {\bibfnamefont {S.}~\bibnamefont
  {Pai}}, \bibinfo {author} {\bibfnamefont {B.}~\bibnamefont {Bartlett}},
  \bibinfo {author} {\bibfnamefont {O.}~\bibnamefont {Solgaard}},\ and\
  \bibinfo {author} {\bibfnamefont {D.~A.}\ \bibnamefont {Miller}},\ }\bibfield
   {title} {\bibinfo {title} {{Matrix optimization on universal unitary
  photonic devices}},\ }\href
  {https://doi.org/10.1103/PhysRevApplied.11.064044} {\bibfield  {journal}
  {\bibinfo  {journal} {Physical Review Applied}\ }\textbf {\bibinfo {volume}
  {11}},\ \bibinfo {pages} {064044} (\bibinfo {year} {2019})}\BibitemShut
  {NoStop}%
\bibitem [{\citenamefont {Bandyopadhyay}\ \emph {et~al.}(2021)\citenamefont
  {Bandyopadhyay}, \citenamefont {Hamerly},\ and\ \citenamefont
  {Englund}}]{hec_1}%
  \BibitemOpen
  \bibfield  {author} {\bibinfo {author} {\bibfnamefont {S.}~\bibnamefont
  {Bandyopadhyay}}, \bibinfo {author} {\bibfnamefont {R.}~\bibnamefont
  {Hamerly}},\ and\ \bibinfo {author} {\bibfnamefont {D.}~\bibnamefont
  {Englund}},\ }\bibfield  {title} {\bibinfo {title} {{Hardware error
  correction for programmable photonics}},\ }\href
  {https://doi.org/10.1364/OPTICA.424052} {\bibfield  {journal} {\bibinfo
  {journal} {Optica}\ }\textbf {\bibinfo {volume} {8}},\ \bibinfo {pages}
  {1247} (\bibinfo {year} {2021})}\BibitemShut {NoStop}%
\bibitem [{\citenamefont {Kumar}\ \emph {et~al.}(2021)\citenamefont {Kumar},
  \citenamefont {Neuhaus}, \citenamefont {Helt}, \citenamefont {Qi},
  \citenamefont {Morrison}, \citenamefont {Mahler},\ and\ \citenamefont
  {Dhand}}]{imperfect_mesh_4}%
  \BibitemOpen
  \bibfield  {author} {\bibinfo {author} {\bibfnamefont {S.~P.}\ \bibnamefont
  {Kumar}}, \bibinfo {author} {\bibfnamefont {L.}~\bibnamefont {Neuhaus}},
  \bibinfo {author} {\bibfnamefont {L.~G.}\ \bibnamefont {Helt}}, \bibinfo
  {author} {\bibfnamefont {H.}~\bibnamefont {Qi}}, \bibinfo {author}
  {\bibfnamefont {B.}~\bibnamefont {Morrison}}, \bibinfo {author}
  {\bibfnamefont {D.~H.}\ \bibnamefont {Mahler}},\ and\ \bibinfo {author}
  {\bibfnamefont {I.}~\bibnamefont {Dhand}},\ }\bibfield  {title} {\bibinfo
  {title} {{Mitigating linear optics imperfections via port allocation and
  compilation}},\ }\bibfield  {journal} {\bibinfo  {journal} {arXiv preprint
  arXiv:2103.03183}\ }\href {https://doi.org/arXiv:2103.03183}
  {arXiv:2103.03183} (\bibinfo {year} {2021})\BibitemShut {NoStop}%
\bibitem [{\citenamefont {Miller}(2017)}]{imperfect_mesh_5}%
  \BibitemOpen
  \bibfield  {author} {\bibinfo {author} {\bibfnamefont {D.~A.}\ \bibnamefont
  {Miller}},\ }\bibfield  {title} {\bibinfo {title} {{Setting up meshes of
  interferometers--reversed local light interference method}},\ }\href
  {https://doi.org/10.1364/OE.25.029233} {\bibfield  {journal} {\bibinfo
  {journal} {Optics Express}\ }\textbf {\bibinfo {volume} {25}},\ \bibinfo
  {pages} {29233} (\bibinfo {year} {2017})}\BibitemShut {NoStop}%
\bibitem [{\citenamefont {Hamerly}\ \emph
  {et~al.}(2022{\natexlab{a}})\citenamefont {Hamerly}, \citenamefont
  {Bandyopadhyay},\ and\ \citenamefont {Englund}}]{self_con_1}%
  \BibitemOpen
  \bibfield  {author} {\bibinfo {author} {\bibfnamefont {R.}~\bibnamefont
  {Hamerly}}, \bibinfo {author} {\bibfnamefont {S.}~\bibnamefont
  {Bandyopadhyay}},\ and\ \bibinfo {author} {\bibfnamefont {D.}~\bibnamefont
  {Englund}},\ }\bibfield  {title} {\bibinfo {title} {{Stability of
  self-configuring large multiport interferometers}},\ }\href
  {https://doi.org/10.1103/PhysRevApplied.18.024018} {\bibfield  {journal}
  {\bibinfo  {journal} {Physical Review Applied}\ }\textbf {\bibinfo {volume}
  {18}},\ \bibinfo {pages} {024018} (\bibinfo {year}
  {2022}{\natexlab{a}})}\BibitemShut {NoStop}%
\bibitem [{\citenamefont {Hamerly}\ \emph
  {et~al.}(2022{\natexlab{b}})\citenamefont {Hamerly}, \citenamefont
  {Bandyopadhyay},\ and\ \citenamefont {Englund}}]{self_con_2}%
  \BibitemOpen
  \bibfield  {author} {\bibinfo {author} {\bibfnamefont {R.}~\bibnamefont
  {Hamerly}}, \bibinfo {author} {\bibfnamefont {S.}~\bibnamefont
  {Bandyopadhyay}},\ and\ \bibinfo {author} {\bibfnamefont {D.}~\bibnamefont
  {Englund}},\ }\bibfield  {title} {\bibinfo {title} {{Accurate
  self-configuration of rectangular multiport interferometers}},\ }\href
  {https://doi.org/10.1103/PhysRevApplied.18.024019} {\bibfield  {journal}
  {\bibinfo  {journal} {Physical Review Applied}\ }\textbf {\bibinfo {volume}
  {18}},\ \bibinfo {pages} {024019} (\bibinfo {year}
  {2022}{\natexlab{b}})}\BibitemShut {NoStop}%
\bibitem [{\citenamefont {Pai}\ \emph {et~al.}(2020)\citenamefont {Pai},
  \citenamefont {Williamson}, \citenamefont {Hughes}, \citenamefont {Minkov},
  \citenamefont {Solgaard}, \citenamefont {Fan},\ and\ \citenamefont
  {Miller}}]{NN_arch_3}%
  \BibitemOpen
  \bibfield  {author} {\bibinfo {author} {\bibfnamefont {S.}~\bibnamefont
  {Pai}}, \bibinfo {author} {\bibfnamefont {I.~A.}\ \bibnamefont {Williamson}},
  \bibinfo {author} {\bibfnamefont {T.~W.}\ \bibnamefont {Hughes}}, \bibinfo
  {author} {\bibfnamefont {M.}~\bibnamefont {Minkov}}, \bibinfo {author}
  {\bibfnamefont {O.}~\bibnamefont {Solgaard}}, \bibinfo {author}
  {\bibfnamefont {S.}~\bibnamefont {Fan}},\ and\ \bibinfo {author}
  {\bibfnamefont {D.~A.}\ \bibnamefont {Miller}},\ }\bibfield  {title}
  {\bibinfo {title} {{Parallel programming of an arbitrary feedforward photonic
  network}},\ }\href {https://doi.org/10.1109/JSTQE.2020.2997849} {\bibfield
  {journal} {\bibinfo  {journal} {IEEE Journal of Selected Topics in Quantum
  Electronics}\ }\textbf {\bibinfo {volume} {26}},\ \bibinfo {pages} {1}
  (\bibinfo {year} {2020})}\BibitemShut {NoStop}%
\bibitem [{\citenamefont {Hamerly}\ \emph {et~al.}(2021)\citenamefont
  {Hamerly}, \citenamefont {Bandyopadhyay},\ and\ \citenamefont
  {Englund}}]{Inf}%
  \BibitemOpen
  \bibfield  {author} {\bibinfo {author} {\bibfnamefont {R.}~\bibnamefont
  {Hamerly}}, \bibinfo {author} {\bibfnamefont {S.}~\bibnamefont
  {Bandyopadhyay}},\ and\ \bibinfo {author} {\bibfnamefont {D.}~\bibnamefont
  {Englund}},\ }\bibfield  {title} {\bibinfo {title} {Infinitely scalable
  multiport interferometers},\ }\bibfield  {journal} {\bibinfo  {journal}
  {arXiv preprint arXiv:2109.05367}\ }\href {https://doi.org/arXiv:2109.05367}
  {arXiv:2109.05367} (\bibinfo {year} {2021})\BibitemShut {NoStop}%
\bibitem [{\citenamefont {Suzuki}\ \emph {et~al.}(2015)\citenamefont {Suzuki},
  \citenamefont {Cong}, \citenamefont {Tanizawa}, \citenamefont {Kim},
  \citenamefont {Ikeda}, \citenamefont {Namiki},\ and\ \citenamefont
  {Kawashima}}]{suzuki2015ultra}%
  \BibitemOpen
  \bibfield  {author} {\bibinfo {author} {\bibfnamefont {K.}~\bibnamefont
  {Suzuki}}, \bibinfo {author} {\bibfnamefont {G.}~\bibnamefont {Cong}},
  \bibinfo {author} {\bibfnamefont {K.}~\bibnamefont {Tanizawa}}, \bibinfo
  {author} {\bibfnamefont {S.-H.}\ \bibnamefont {Kim}}, \bibinfo {author}
  {\bibfnamefont {K.}~\bibnamefont {Ikeda}}, \bibinfo {author} {\bibfnamefont
  {S.}~\bibnamefont {Namiki}},\ and\ \bibinfo {author} {\bibfnamefont
  {H.}~\bibnamefont {Kawashima}},\ }\bibfield  {title} {\bibinfo {title}
  {{Ultra-high-extinction-ratio 2$\times$ 2 silicon optical switch with
  variable splitter}},\ }\href {https://doi.org/10.1364/OE.23.009086}
  {\bibfield  {journal} {\bibinfo  {journal} {Optics Express}\ }\textbf
  {\bibinfo {volume} {23}},\ \bibinfo {pages} {9086} (\bibinfo {year}
  {2015})}\BibitemShut {NoStop}%
\bibitem [{\citenamefont {Miller}(2015)}]{miller2015perfect}%
  \BibitemOpen
  \bibfield  {author} {\bibinfo {author} {\bibfnamefont {D.~A.}\ \bibnamefont
  {Miller}},\ }\bibfield  {title} {\bibinfo {title} {{Perfect optics with
  imperfect components}},\ }\href {https://doi.org/10.1364/OPTICA.2.000747}
  {\bibfield  {journal} {\bibinfo  {journal} {Optica}\ }\textbf {\bibinfo
  {volume} {2}},\ \bibinfo {pages} {747} (\bibinfo {year} {2015})}\BibitemShut
  {NoStop}%
\bibitem [{\citenamefont {Polcari}(2018)}]{butterfly_decomp}%
  \BibitemOpen
  \bibfield  {author} {\bibinfo {author} {\bibfnamefont {J.}~\bibnamefont
  {Polcari}},\ }\bibfield  {title} {\bibinfo {title} {{Generalizing the
  Butterfly Structure of the FFT}},\ }in\ \href
  {https://doi.org/10.1007/978-3-319-95117-1_3} {\emph {\bibinfo {booktitle}
  {Advanced Research in Naval Engineering}}}\ (\bibinfo  {publisher}
  {Springer},\ \bibinfo {year} {2018})\ pp.\ \bibinfo {pages}
  {35--52}\BibitemShut {NoStop}%
\bibitem [{\citenamefont {Reck}\ \emph {et~al.}(1994)\citenamefont {Reck},
  \citenamefont {Zeilinger}, \citenamefont {Bernstein},\ and\ \citenamefont
  {Bertani}}]{reck}%
  \BibitemOpen
  \bibfield  {author} {\bibinfo {author} {\bibfnamefont {M.}~\bibnamefont
  {Reck}}, \bibinfo {author} {\bibfnamefont {A.}~\bibnamefont {Zeilinger}},
  \bibinfo {author} {\bibfnamefont {H.~J.}\ \bibnamefont {Bernstein}},\ and\
  \bibinfo {author} {\bibfnamefont {P.}~\bibnamefont {Bertani}},\ }\bibfield
  {title} {\bibinfo {title} {{Experimental realization of any discrete unitary
  operator}},\ }\href {https://doi.org/10.1103/PhysRevLett.73.58} {\bibfield
  {journal} {\bibinfo  {journal} {Physical Review Letters}\ }\textbf {\bibinfo
  {volume} {73}},\ \bibinfo {pages} {58} (\bibinfo {year} {1994})}\BibitemShut
  {NoStop}%
\bibitem [{\citenamefont {Clements}\ \emph {et~al.}(2016)\citenamefont
  {Clements}, \citenamefont {Humphreys}, \citenamefont {Metcalf}, \citenamefont
  {Kolthammer},\ and\ \citenamefont {Walmsley}}]{clements}%
  \BibitemOpen
  \bibfield  {author} {\bibinfo {author} {\bibfnamefont {W.~R.}\ \bibnamefont
  {Clements}}, \bibinfo {author} {\bibfnamefont {P.~C.}\ \bibnamefont
  {Humphreys}}, \bibinfo {author} {\bibfnamefont {B.~J.}\ \bibnamefont
  {Metcalf}}, \bibinfo {author} {\bibfnamefont {W.~S.}\ \bibnamefont
  {Kolthammer}},\ and\ \bibinfo {author} {\bibfnamefont {I.~A.}\ \bibnamefont
  {Walmsley}},\ }\bibfield  {title} {\bibinfo {title} {{Optimal design for
  universal multiport interferometers}},\ }\href
  {https://doi.org/10.1364/OPTICA.3.001460} {\bibfield  {journal} {\bibinfo
  {journal} {Optica}\ }\textbf {\bibinfo {volume} {3}},\ \bibinfo {pages}
  {1460} (\bibinfo {year} {2016})}\BibitemShut {NoStop}%
\bibitem [{\citenamefont {Cooley}\ and\ \citenamefont
  {Tukey}(1965)}]{butterfly_og}%
  \BibitemOpen
  \bibfield  {author} {\bibinfo {author} {\bibfnamefont {J.~W.}\ \bibnamefont
  {Cooley}}\ and\ \bibinfo {author} {\bibfnamefont {J.~W.}\ \bibnamefont
  {Tukey}},\ }\bibfield  {title} {\bibinfo {title} {{An algorithm for the
  machine calculation of complex Fourier series}},\ }\href
  {https://doi.org/10.1090/S0025-5718-1965-0178586-1} {\bibfield  {journal}
  {\bibinfo  {journal} {Mathematics of computation}\ }\textbf {\bibinfo
  {volume} {19}},\ \bibinfo {pages} {297} (\bibinfo {year} {1965})}\BibitemShut
  {NoStop}%
\bibitem [{\citenamefont {Flamini}\ \emph {et~al.}(2017)\citenamefont
  {Flamini}, \citenamefont {Spagnolo}, \citenamefont {Viggianiello},
  \citenamefont {Crespi}, \citenamefont {Osellame},\ and\ \citenamefont
  {Sciarrino}}]{FFT_analog_1}%
  \BibitemOpen
  \bibfield  {author} {\bibinfo {author} {\bibfnamefont {F.}~\bibnamefont
  {Flamini}}, \bibinfo {author} {\bibfnamefont {N.}~\bibnamefont {Spagnolo}},
  \bibinfo {author} {\bibfnamefont {N.}~\bibnamefont {Viggianiello}}, \bibinfo
  {author} {\bibfnamefont {A.}~\bibnamefont {Crespi}}, \bibinfo {author}
  {\bibfnamefont {R.}~\bibnamefont {Osellame}},\ and\ \bibinfo {author}
  {\bibfnamefont {F.}~\bibnamefont {Sciarrino}},\ }\bibfield  {title} {\bibinfo
  {title} {{Benchmarking integrated linear-optical architectures for quantum
  information processing}},\ }\href
  {https://doi.org/10.1038/s41598-017-15174-2} {\bibfield  {journal} {\bibinfo
  {journal} {Scientific Reports}\ }\textbf {\bibinfo {volume} {7}},\ \bibinfo
  {pages} {1} (\bibinfo {year} {2017})}\BibitemShut {NoStop}%
\bibitem [{\citenamefont {Gu}\ \emph {et~al.}(2020)\citenamefont {Gu},
  \citenamefont {Zhao}, \citenamefont {Feng}, \citenamefont {Liu},
  \citenamefont {Chen},\ and\ \citenamefont {Pan}}]{FFT_analog_2}%
  \BibitemOpen
  \bibfield  {author} {\bibinfo {author} {\bibfnamefont {J.}~\bibnamefont
  {Gu}}, \bibinfo {author} {\bibfnamefont {Z.}~\bibnamefont {Zhao}}, \bibinfo
  {author} {\bibfnamefont {C.}~\bibnamefont {Feng}}, \bibinfo {author}
  {\bibfnamefont {M.}~\bibnamefont {Liu}}, \bibinfo {author} {\bibfnamefont
  {R.~T.}\ \bibnamefont {Chen}},\ and\ \bibinfo {author} {\bibfnamefont
  {D.~Z.}\ \bibnamefont {Pan}},\ }\bibfield  {title} {\bibinfo {title}
  {{Towards area-efficient optical neural networks: an FFT-based
  architecture}},\ }in\ \href
  {https://doi.org/10.1109/ASP-DAC47756.2020.9045156} {\emph {\bibinfo
  {booktitle} {2020 25th Asia and South Pacific Design Automation Conference
  (ASP-DAC)}}}\ (\bibinfo {organization} {IEEE},\ \bibinfo {year} {2020})\ pp.\
  \bibinfo {pages} {476--481}\BibitemShut {NoStop}%
\bibitem [{\citenamefont {Hinton}\ \emph {et~al.}(2012)\citenamefont {Hinton},
  \citenamefont {Srivastava}, \citenamefont {Krizhevsky}, \citenamefont
  {Sutskever},\ and\ \citenamefont {Salakhutdinov}}]{conventional_prune_1}%
  \BibitemOpen
  \bibfield  {author} {\bibinfo {author} {\bibfnamefont {G.~E.}\ \bibnamefont
  {Hinton}}, \bibinfo {author} {\bibfnamefont {N.}~\bibnamefont {Srivastava}},
  \bibinfo {author} {\bibfnamefont {A.}~\bibnamefont {Krizhevsky}}, \bibinfo
  {author} {\bibfnamefont {I.}~\bibnamefont {Sutskever}},\ and\ \bibinfo
  {author} {\bibfnamefont {R.~R.}\ \bibnamefont {Salakhutdinov}},\ }\bibfield
  {title} {\bibinfo {title} {{Improving neural networks by preventing
  co-adaptation of feature detectors}},\ }\bibfield  {journal} {\bibinfo
  {journal} {arXiv preprint arXiv:1207.0580}\ }\href
  {https://doi.org/arXiv:1207.0580} {arXiv:1207.0580} (\bibinfo {year}
  {2012})\BibitemShut {NoStop}%
\bibitem [{\citenamefont {Tanaka}\ \emph {et~al.}(2020)\citenamefont {Tanaka},
  \citenamefont {Kunin}, \citenamefont {Yamins},\ and\ \citenamefont
  {Ganguli}}]{conventional_prune_2}%
  \BibitemOpen
  \bibfield  {author} {\bibinfo {author} {\bibfnamefont {H.}~\bibnamefont
  {Tanaka}}, \bibinfo {author} {\bibfnamefont {D.}~\bibnamefont {Kunin}},
  \bibinfo {author} {\bibfnamefont {D.~L.}\ \bibnamefont {Yamins}},\ and\
  \bibinfo {author} {\bibfnamefont {S.}~\bibnamefont {Ganguli}},\ }\bibfield
  {title} {\bibinfo {title} {Pruning neural networks without any data by
  iteratively conserving synaptic flow},\ }\href
  {https://doi.org/arXiv:2006.05467} {\bibfield  {journal} {\bibinfo  {journal}
  {Advances in Neural Information Processing Systems}\ }\textbf {\bibinfo
  {volume} {33}},\ \bibinfo {pages} {6377} (\bibinfo {year}
  {2020})}\BibitemShut {NoStop}%
\bibitem [{\citenamefont {Lee}\ and\ \citenamefont
  {Park}(2007)}]{binary_tree_decomp}%
  \BibitemOpen
  \bibfield  {author} {\bibinfo {author} {\bibfnamefont {H.-Y.}\ \bibnamefont
  {Lee}}\ and\ \bibinfo {author} {\bibfnamefont {I.-C.}\ \bibnamefont {Park}},\
  }\bibfield  {title} {\bibinfo {title} {{Balanced binary-tree decomposition
  for area-efficient pipelined FFT processing}},\ }\href
  {https://doi.org/10.1109/TCSI.2006.888764} {\bibfield  {journal} {\bibinfo
  {journal} {IEEE Transactions on Circuits and Systems I: Regular Papers}\
  }\textbf {\bibinfo {volume} {54}},\ \bibinfo {pages} {889} (\bibinfo {year}
  {2007})}\BibitemShut {NoStop}%
\bibitem [{\citenamefont {Tung}(2003)}]{Haar_sampling}%
  \BibitemOpen
  \bibfield  {author} {\bibinfo {author} {\bibfnamefont {W.-K.}\ \bibnamefont
  {Tung}},\ }\href {https://doi.org/10.1142/0097} {\bibinfo {title} {{Group
  Theory in Physics: An Introduction to Symmetry Principles, Group
  Representations, and Special Functions in Classical and Quantum Physics}}}
  (\bibinfo {year} {2003})\BibitemShut {NoStop}%
\bibitem [{\citenamefont {Russell}\ \emph {et~al.}(2017)\citenamefont
  {Russell}, \citenamefont {Chakhmakhchyan}, \citenamefont {O’Brien},\ and\
  \citenamefont {Laing}}]{MZI_rank}%
  \BibitemOpen
  \bibfield  {author} {\bibinfo {author} {\bibfnamefont {N.~J.}\ \bibnamefont
  {Russell}}, \bibinfo {author} {\bibfnamefont {L.}~\bibnamefont
  {Chakhmakhchyan}}, \bibinfo {author} {\bibfnamefont {J.~L.}\ \bibnamefont
  {O’Brien}},\ and\ \bibinfo {author} {\bibfnamefont {A.}~\bibnamefont
  {Laing}},\ }\bibfield  {title} {\bibinfo {title} {{Direct dialling of Haar
  random unitary matrices}},\ }\href {https://doi.org/10.1088/1367-2630/aa60ed}
  {\bibfield  {journal} {\bibinfo  {journal} {New Journal of Physics}\ }\textbf
  {\bibinfo {volume} {19}},\ \bibinfo {pages} {033007} (\bibinfo {year}
  {2017})}\BibitemShut {NoStop}%
\bibitem [{\citenamefont {Williamson}\ \emph {et~al.}(2019)\citenamefont
  {Williamson}, \citenamefont {Hughes}, \citenamefont {Minkov}, \citenamefont
  {Bartlett}, \citenamefont {Pai},\ and\ \citenamefont {Fan}}]{MZI_NL}%
  \BibitemOpen
  \bibfield  {author} {\bibinfo {author} {\bibfnamefont {I.~A.}\ \bibnamefont
  {Williamson}}, \bibinfo {author} {\bibfnamefont {T.~W.}\ \bibnamefont
  {Hughes}}, \bibinfo {author} {\bibfnamefont {M.}~\bibnamefont {Minkov}},
  \bibinfo {author} {\bibfnamefont {B.}~\bibnamefont {Bartlett}}, \bibinfo
  {author} {\bibfnamefont {S.}~\bibnamefont {Pai}},\ and\ \bibinfo {author}
  {\bibfnamefont {S.}~\bibnamefont {Fan}},\ }\bibfield  {title} {\bibinfo
  {title} {{Reprogrammable electro-optic nonlinear activation functions for
  optical neural networks}},\ }\href
  {https://doi.org/10.1109/JSTQE.2019.2930455} {\bibfield  {journal} {\bibinfo
  {journal} {IEEE Journal of Selected Topics in Quantum Electronics}\ }\textbf
  {\bibinfo {volume} {26}},\ \bibinfo {pages} {1} (\bibinfo {year}
  {2019})}\BibitemShut {NoStop}%
\bibitem [{\citenamefont {Hamerly}(2021)}]{meshes_ryan}%
  \BibitemOpen
  \bibfield  {author} {\bibinfo {author} {\bibfnamefont {R.}~\bibnamefont
  {Hamerly}},\ }\href@noop {} {\bibinfo {title} {{Meshes: Tools for modeling
  photonic beamsplitter mesh networks}}},\ \bibinfo {howpublished}
  {\url{https://github. com/QPG-MIT/meshes}} (\bibinfo {year}
  {2021})\BibitemShut {NoStop}%
\bibitem [{\citenamefont {LeCun}(1998)}]{MNIST}%
  \BibitemOpen
  \bibfield  {author} {\bibinfo {author} {\bibfnamefont {Y.}~\bibnamefont
  {LeCun}},\ }\href@noop {} {\bibinfo {title} {{The MNIST database of
  handwritten digits}}},\ \bibinfo {howpublished} {\url{http://yann. lecun.
  com/exdb/mnist/}} (\bibinfo {year} {1998})\BibitemShut {NoStop}%
\bibitem [{\citenamefont {Mikkelsen}\ \emph {et~al.}(2014)\citenamefont
  {Mikkelsen}, \citenamefont {Sacher},\ and\ \citenamefont
  {Poon}}]{process_err}%
  \BibitemOpen
  \bibfield  {author} {\bibinfo {author} {\bibfnamefont {J.~C.}\ \bibnamefont
  {Mikkelsen}}, \bibinfo {author} {\bibfnamefont {W.~D.}\ \bibnamefont
  {Sacher}},\ and\ \bibinfo {author} {\bibfnamefont {J.~K.}\ \bibnamefont
  {Poon}},\ }\bibfield  {title} {\bibinfo {title} {{Dimensional variation
  tolerant silicon-on-insulator directional couplers}},\ }\href
  {https://doi.org/10.1364/OE.22.003145} {\bibfield  {journal} {\bibinfo
  {journal} {Optics Express}\ }\textbf {\bibinfo {volume} {22}},\ \bibinfo
  {pages} {3145} (\bibinfo {year} {2014})}\BibitemShut {NoStop}%
\bibitem [{\citenamefont {Hamerly}\ \emph {et~al.}(2019)\citenamefont
  {Hamerly}, \citenamefont {Bernstein}, \citenamefont {Sludds}, \citenamefont
  {Solja{\v{c}}i{\'c}},\ and\ \citenamefont {Englund}}]{Opt_DNN_Ryan}%
  \BibitemOpen
  \bibfield  {author} {\bibinfo {author} {\bibfnamefont {R.}~\bibnamefont
  {Hamerly}}, \bibinfo {author} {\bibfnamefont {L.}~\bibnamefont {Bernstein}},
  \bibinfo {author} {\bibfnamefont {A.}~\bibnamefont {Sludds}}, \bibinfo
  {author} {\bibfnamefont {M.}~\bibnamefont {Solja{\v{c}}i{\'c}}},\ and\
  \bibinfo {author} {\bibfnamefont {D.}~\bibnamefont {Englund}},\ }\bibfield
  {title} {\bibinfo {title} {{Large-scale optical neural networks based on
  photoelectric multiplication}},\ }\href
  {https://doi.org/10.1103/PhysRevX.9.021032} {\bibfield  {journal} {\bibinfo
  {journal} {Physical Review X}\ }\textbf {\bibinfo {volume} {9}},\ \bibinfo
  {pages} {021032} (\bibinfo {year} {2019})}\BibitemShut {NoStop}%
\bibitem [{\citenamefont {Vadlamani}\ \emph {et~al.}(2022)\citenamefont
  {Vadlamani}, \citenamefont {Englund},\ and\ \citenamefont
  {Hamerly}}]{Sri_paper}%
  \BibitemOpen
  \bibfield  {author} {\bibinfo {author} {\bibfnamefont {S.~K.}\ \bibnamefont
  {Vadlamani}}, \bibinfo {author} {\bibfnamefont {D.}~\bibnamefont {Englund}},\
  and\ \bibinfo {author} {\bibfnamefont {R.}~\bibnamefont {Hamerly}},\
  }\href@noop {} {\bibinfo {title} {{Transferable Learning on Analog
  Hardware}}},\ \bibinfo {howpublished} {\textit{In preparation}} (\bibinfo
  {year} {2022})\BibitemShut {NoStop}%
\bibitem [{\citenamefont {Jones}\ \emph {et~al.}(2013)\citenamefont {Jones},
  \citenamefont {DeRose}, \citenamefont {Lentine}, \citenamefont {Trotter},
  \citenamefont {Starbuck},\ and\ \citenamefont {Norwood}}]{3d_cross_1}%
  \BibitemOpen
  \bibfield  {author} {\bibinfo {author} {\bibfnamefont {A.~M.}\ \bibnamefont
  {Jones}}, \bibinfo {author} {\bibfnamefont {C.~T.}\ \bibnamefont {DeRose}},
  \bibinfo {author} {\bibfnamefont {A.~L.}\ \bibnamefont {Lentine}}, \bibinfo
  {author} {\bibfnamefont {D.~C.}\ \bibnamefont {Trotter}}, \bibinfo {author}
  {\bibfnamefont {A.~L.}\ \bibnamefont {Starbuck}},\ and\ \bibinfo {author}
  {\bibfnamefont {R.~A.}\ \bibnamefont {Norwood}},\ }\bibfield  {title}
  {\bibinfo {title} {{Ultra-low crosstalk, CMOS compatible waveguide crossings
  for densely integrated photonic interconnection networks}},\ }\href
  {https://doi.org/10.1364/OE.21.012002} {\bibfield  {journal} {\bibinfo
  {journal} {Optics Express}\ }\textbf {\bibinfo {volume} {21}},\ \bibinfo
  {pages} {12002} (\bibinfo {year} {2013})}\BibitemShut {NoStop}%
\bibitem [{\citenamefont {Sacher}\ \emph {et~al.}(2018)\citenamefont {Sacher},
  \citenamefont {Mikkelsen}, \citenamefont {Huang}, \citenamefont {Mak},
  \citenamefont {Yong}, \citenamefont {Luo}, \citenamefont {Li}, \citenamefont
  {Dumais}, \citenamefont {Jiang}, \citenamefont {Goodwill} \emph
  {et~al.}}]{3d_cross_2}%
  \BibitemOpen
  \bibfield  {author} {\bibinfo {author} {\bibfnamefont {W.~D.}\ \bibnamefont
  {Sacher}}, \bibinfo {author} {\bibfnamefont {J.~C.}\ \bibnamefont
  {Mikkelsen}}, \bibinfo {author} {\bibfnamefont {Y.}~\bibnamefont {Huang}},
  \bibinfo {author} {\bibfnamefont {J.~C.}\ \bibnamefont {Mak}}, \bibinfo
  {author} {\bibfnamefont {Z.}~\bibnamefont {Yong}}, \bibinfo {author}
  {\bibfnamefont {X.}~\bibnamefont {Luo}}, \bibinfo {author} {\bibfnamefont
  {Y.}~\bibnamefont {Li}}, \bibinfo {author} {\bibfnamefont {P.}~\bibnamefont
  {Dumais}}, \bibinfo {author} {\bibfnamefont {J.}~\bibnamefont {Jiang}},
  \bibinfo {author} {\bibfnamefont {D.}~\bibnamefont {Goodwill}}, \emph
  {et~al.},\ }\bibfield  {title} {\bibinfo {title} {{Monolithically integrated
  multilayer silicon nitride-on-silicon waveguide platforms for 3-D photonic
  circuits and devices}},\ }\href {https://doi.org/10.1109/JPROC.2018.2860994}
  {\bibfield  {journal} {\bibinfo  {journal} {Proceedings of the IEEE}\
  }\textbf {\bibinfo {volume} {106}},\ \bibinfo {pages} {2232} (\bibinfo {year}
  {2018})}\BibitemShut {NoStop}%
\bibitem [{\citenamefont {Lindenmann}\ \emph {et~al.}(2012)\citenamefont
  {Lindenmann}, \citenamefont {Balthasar}, \citenamefont {Hillerkuss},
  \citenamefont {Schmogrow}, \citenamefont {Jordan}, \citenamefont {Leuthold},
  \citenamefont {Freude},\ and\ \citenamefont {Koos}}]{3d_cross_polymer_1}%
  \BibitemOpen
  \bibfield  {author} {\bibinfo {author} {\bibfnamefont {N.}~\bibnamefont
  {Lindenmann}}, \bibinfo {author} {\bibfnamefont {G.}~\bibnamefont
  {Balthasar}}, \bibinfo {author} {\bibfnamefont {D.}~\bibnamefont
  {Hillerkuss}}, \bibinfo {author} {\bibfnamefont {R.}~\bibnamefont
  {Schmogrow}}, \bibinfo {author} {\bibfnamefont {M.}~\bibnamefont {Jordan}},
  \bibinfo {author} {\bibfnamefont {J.}~\bibnamefont {Leuthold}}, \bibinfo
  {author} {\bibfnamefont {W.}~\bibnamefont {Freude}},\ and\ \bibinfo {author}
  {\bibfnamefont {C.}~\bibnamefont {Koos}},\ }\bibfield  {title} {\bibinfo
  {title} {{Photonic wire bonding: a novel concept for chip-scale
  interconnects}},\ }\href {https://doi.org/10.1364/OE.20.017667} {\bibfield
  {journal} {\bibinfo  {journal} {Optics Express}\ }\textbf {\bibinfo {volume}
  {20}},\ \bibinfo {pages} {17667} (\bibinfo {year} {2012})}\BibitemShut
  {NoStop}%
\bibitem [{\citenamefont {Billah}\ \emph {et~al.}(2018)\citenamefont {Billah},
  \citenamefont {Blaicher}, \citenamefont {Hoose}, \citenamefont {Dietrich},
  \citenamefont {Marin-Palomo}, \citenamefont {Lindenmann}, \citenamefont
  {Nesic}, \citenamefont {Hofmann}, \citenamefont {Troppenz}, \citenamefont
  {Moehrle} \emph {et~al.}}]{3d_cross_polymer_2}%
  \BibitemOpen
  \bibfield  {author} {\bibinfo {author} {\bibfnamefont {M.~R.}\ \bibnamefont
  {Billah}}, \bibinfo {author} {\bibfnamefont {M.}~\bibnamefont {Blaicher}},
  \bibinfo {author} {\bibfnamefont {T.}~\bibnamefont {Hoose}}, \bibinfo
  {author} {\bibfnamefont {P.-I.}\ \bibnamefont {Dietrich}}, \bibinfo {author}
  {\bibfnamefont {P.}~\bibnamefont {Marin-Palomo}}, \bibinfo {author}
  {\bibfnamefont {N.}~\bibnamefont {Lindenmann}}, \bibinfo {author}
  {\bibfnamefont {A.}~\bibnamefont {Nesic}}, \bibinfo {author} {\bibfnamefont
  {A.}~\bibnamefont {Hofmann}}, \bibinfo {author} {\bibfnamefont
  {U.}~\bibnamefont {Troppenz}}, \bibinfo {author} {\bibfnamefont
  {M.}~\bibnamefont {Moehrle}}, \emph {et~al.},\ }\bibfield  {title} {\bibinfo
  {title} {{Hybrid integration of silicon photonics circuits and InP lasers by
  photonic wire bonding}},\ }\href {https://doi.org/10.1364/OPTICA.5.000876}
  {\bibfield  {journal} {\bibinfo  {journal} {Optica}\ }\textbf {\bibinfo
  {volume} {5}},\ \bibinfo {pages} {876} (\bibinfo {year} {2018})}\BibitemShut
  {NoStop}%
\bibitem [{\citenamefont {Mennea}\ \emph {et~al.}(2018)\citenamefont {Mennea},
  \citenamefont {Clements}, \citenamefont {Smith}, \citenamefont {Gates},
  \citenamefont {Metcalf}, \citenamefont {Bannerman}, \citenamefont {Burgwal},
  \citenamefont {Renema}, \citenamefont {Kolthammer}, \citenamefont {Walmsley}
  \emph {et~al.}}]{3d_cross_glass_1}%
  \BibitemOpen
  \bibfield  {author} {\bibinfo {author} {\bibfnamefont {P.~L.}\ \bibnamefont
  {Mennea}}, \bibinfo {author} {\bibfnamefont {W.~R.}\ \bibnamefont
  {Clements}}, \bibinfo {author} {\bibfnamefont {D.~H.}\ \bibnamefont {Smith}},
  \bibinfo {author} {\bibfnamefont {J.~C.}\ \bibnamefont {Gates}}, \bibinfo
  {author} {\bibfnamefont {B.~J.}\ \bibnamefont {Metcalf}}, \bibinfo {author}
  {\bibfnamefont {R.~H.}\ \bibnamefont {Bannerman}}, \bibinfo {author}
  {\bibfnamefont {R.}~\bibnamefont {Burgwal}}, \bibinfo {author} {\bibfnamefont
  {J.~J.}\ \bibnamefont {Renema}}, \bibinfo {author} {\bibfnamefont {W.~S.}\
  \bibnamefont {Kolthammer}}, \bibinfo {author} {\bibfnamefont {I.~A.}\
  \bibnamefont {Walmsley}}, \emph {et~al.},\ }\bibfield  {title} {\bibinfo
  {title} {{Modular linear optical circuits}},\ }\href
  {https://doi.org/10.1364/OPTICA.5.001087} {\bibfield  {journal} {\bibinfo
  {journal} {Optica}\ }\textbf {\bibinfo {volume} {5}},\ \bibinfo {pages}
  {1087} (\bibinfo {year} {2018})}\BibitemShut {NoStop}%
\bibitem [{\citenamefont {Szameit}\ and\ \citenamefont
  {Nolte}(2010)}]{3d_cross_glass_2}%
  \BibitemOpen
  \bibfield  {author} {\bibinfo {author} {\bibfnamefont {A.}~\bibnamefont
  {Szameit}}\ and\ \bibinfo {author} {\bibfnamefont {S.}~\bibnamefont
  {Nolte}},\ }\bibfield  {title} {\bibinfo {title} {{Discrete optics in
  femtosecond-laser-written photonic structures}},\ }\href
  {https://doi.org/10.1088/0953-4075/43/16/163001} {\bibfield  {journal}
  {\bibinfo  {journal} {Journal of Physics B: Atomic, Molecular and Optical
  Physics}\ }\textbf {\bibinfo {volume} {43}},\ \bibinfo {pages} {163001}
  (\bibinfo {year} {2010})}\BibitemShut {NoStop}%
\bibitem [{\citenamefont {Bandyopadhyay}\ and\ \citenamefont
  {Englund}(2021)}]{HockeyStick}%
  \BibitemOpen
  \bibfield  {author} {\bibinfo {author} {\bibfnamefont {S.}~\bibnamefont
  {Bandyopadhyay}}\ and\ \bibinfo {author} {\bibfnamefont {D.}~\bibnamefont
  {Englund}},\ }\bibfield  {title} {\bibinfo {title} {Alignment-free photonic
  interconnects},\ }\bibfield  {journal} {\bibinfo  {journal} {arXiv preprint
  arXiv:2110.12851}\ }\href {https://doi.org/arXiv:2110.12851}
  {arXiv:2110.12851} (\bibinfo {year} {2021})\BibitemShut {NoStop}%
\bibitem [{\citenamefont {Friedmann}\ \emph {et~al.}(2013)\citenamefont
  {Friedmann}, \citenamefont {Fr{\'e}maux}, \citenamefont {Schemmel},
  \citenamefont {Gerstner},\ and\ \citenamefont {Meier}}]{DNN_acc_1}%
  \BibitemOpen
  \bibfield  {author} {\bibinfo {author} {\bibfnamefont {S.}~\bibnamefont
  {Friedmann}}, \bibinfo {author} {\bibfnamefont {N.}~\bibnamefont
  {Fr{\'e}maux}}, \bibinfo {author} {\bibfnamefont {J.}~\bibnamefont
  {Schemmel}}, \bibinfo {author} {\bibfnamefont {W.}~\bibnamefont {Gerstner}},\
  and\ \bibinfo {author} {\bibfnamefont {K.}~\bibnamefont {Meier}},\ }\bibfield
   {title} {\bibinfo {title} {{Reward-based learning under hardware
  constraints—using a RISC processor embedded in a neuromorphic substrate}},\
  }\href {https://doi.org/10.3389/fnins.2013.00160} {\bibfield  {journal}
  {\bibinfo  {journal} {Frontiers in Neuroscience}\ }\textbf {\bibinfo {volume}
  {7}},\ \bibinfo {pages} {160} (\bibinfo {year} {2013})}\BibitemShut {NoStop}%
\bibitem [{\citenamefont {Akopyan}\ \emph {et~al.}(2015)\citenamefont
  {Akopyan}, \citenamefont {Sawada}, \citenamefont {Cassidy}, \citenamefont
  {Alvarez-Icaza}, \citenamefont {Arthur}, \citenamefont {Merolla},
  \citenamefont {Imam}, \citenamefont {Nakamura}, \citenamefont {Datta},
  \citenamefont {Nam} \emph {et~al.}}]{DNN_acc_2}%
  \BibitemOpen
  \bibfield  {author} {\bibinfo {author} {\bibfnamefont {F.}~\bibnamefont
  {Akopyan}}, \bibinfo {author} {\bibfnamefont {J.}~\bibnamefont {Sawada}},
  \bibinfo {author} {\bibfnamefont {A.}~\bibnamefont {Cassidy}}, \bibinfo
  {author} {\bibfnamefont {R.}~\bibnamefont {Alvarez-Icaza}}, \bibinfo {author}
  {\bibfnamefont {J.}~\bibnamefont {Arthur}}, \bibinfo {author} {\bibfnamefont
  {P.}~\bibnamefont {Merolla}}, \bibinfo {author} {\bibfnamefont
  {N.}~\bibnamefont {Imam}}, \bibinfo {author} {\bibfnamefont {Y.}~\bibnamefont
  {Nakamura}}, \bibinfo {author} {\bibfnamefont {P.}~\bibnamefont {Datta}},
  \bibinfo {author} {\bibfnamefont {G.-J.}\ \bibnamefont {Nam}}, \emph
  {et~al.},\ }\bibfield  {title} {\bibinfo {title} {{Truenorth: Design and tool
  flow of a 65 mw 1 million neuron programmable neurosynaptic chip}},\ }\href
  {https://doi.org/10.1109/TCAD.2015.2474396} {\bibfield  {journal} {\bibinfo
  {journal} {IEEE transactions on computer-aided design of integrated circuits
  and systems}\ }\textbf {\bibinfo {volume} {34}},\ \bibinfo {pages} {1537}
  (\bibinfo {year} {2015})}\BibitemShut {NoStop}%
\bibitem [{\citenamefont {Jouppi}\ \emph {et~al.}(2017)\citenamefont {Jouppi},
  \citenamefont {Young}, \citenamefont {Patil}, \citenamefont {Patterson},
  \citenamefont {Agrawal}, \citenamefont {Bajwa}, \citenamefont {Bates},
  \citenamefont {Bhatia}, \citenamefont {Boden}, \citenamefont {Borchers} \emph
  {et~al.}}]{DNN_acc_3}%
  \BibitemOpen
  \bibfield  {author} {\bibinfo {author} {\bibfnamefont {N.~P.}\ \bibnamefont
  {Jouppi}}, \bibinfo {author} {\bibfnamefont {C.}~\bibnamefont {Young}},
  \bibinfo {author} {\bibfnamefont {N.}~\bibnamefont {Patil}}, \bibinfo
  {author} {\bibfnamefont {D.}~\bibnamefont {Patterson}}, \bibinfo {author}
  {\bibfnamefont {G.}~\bibnamefont {Agrawal}}, \bibinfo {author} {\bibfnamefont
  {R.}~\bibnamefont {Bajwa}}, \bibinfo {author} {\bibfnamefont
  {S.}~\bibnamefont {Bates}}, \bibinfo {author} {\bibfnamefont
  {S.}~\bibnamefont {Bhatia}}, \bibinfo {author} {\bibfnamefont
  {N.}~\bibnamefont {Boden}}, \bibinfo {author} {\bibfnamefont
  {A.}~\bibnamefont {Borchers}}, \emph {et~al.},\ }\bibfield  {title} {\bibinfo
  {title} {{In-datacenter performance analysis of a tensor processing unit}},\
  }in\ \href {https://doi.org/10.1145/3079856.3080246} {\emph {\bibinfo
  {booktitle} {Proceedings of the 44th annual international symposium on
  computer architecture}}}\ (\bibinfo {year} {2017})\ pp.\ \bibinfo {pages}
  {1--12}\BibitemShut {NoStop}%
\end{thebibliography}%

\end{document}